\documentclass[12pt]{article}

\usepackage{latexsym}
\usepackage{graphics}
\usepackage{epsfig}
\usepackage[mathscr]{eucal}
\usepackage{amscd}
\usepackage{amssymb}
\usepackage{amsmath}
\newcommand{\re}{\mathbb R}

\newcommand{\mmin}{\mathrm {min}}
\newcommand{\mmax}{\mathrm {max}}

\def\be{\begin{displaymath}}
\def\ee{\end{displaymath}}

\newcommand{\causet}{C}
\newcommand{\manifold}{ M}
\newcommand{\spacetime}{( M, g)}

\newcommand{\R}{\mathbb R}
\newcommand{\Z}{\mathbb Z}

\newcommand{\fut}{\mathrm{fut}}
\newcommand{\past}{\mathrm{past}}

\newcommand{\ta}{{ A}}
\newcommand{\tta}{ T(A)}
\newcommand{\ttav}{ T_n(A)}
\newcommand{\ttao}{ T_0(A)}
\usepackage{psfrag}
\usepackage[dvips]{color}

\newcommand{\cO}{\mathcal O}
\newcommand{\bZ}{\mathbb Z}
\newcommand{\cN}{\mathcal N}

\newcommand{\sn}{s_R}
\newcommand{\lam}{\lambda}

\title{Stable Homology as an Indicator of Manifoldlikeness in Causal
  Set Theory}

\author{Seth Major${}^1$, David Rideout${}^2$,  and Sumati Surya${}^3$ \\
${}^1$ Hamilton College, Clinton, NY, USA\\
${}^2$ Perimeter Institute for Theoretical Physics, Waterloo, ON, Canada\\
${}^3$ Raman Research Institute, Bangalore, India}
\begin{document} 
\maketitle
\begin{abstract} 
We present a computational tool that can be used to obtain the
``spatial'' homology groups of a causal set. Localisation in the
causal set is seeded by an inextendible antichain, which is the analog
of a spacelike hypersurface, and a one parameter family of nerve
simplicial complexes is constructed by ``thickening'' this antichain.
The associated homology groups can then be calculated using existing
homology software, and their behaviour studied as a function of the
thickening parameter.  Earlier analytical work showed that for an
inextendible antichain in a causal set which can be approximated by a
globally hyperbolic spacetime region, there is a one parameter
sub-family of these simplicial complexes which are homological to the
continuum, provided the antichain satisfies certain conditions. Using
causal sets that are approximated by a set of 2d spacetimes our
numerical analysis suggests that these conditions are generically
satisfied by inextendible antichains.  In both 2d and 3d simulations,
as the thickening parameter is increased, the continuum homology
groups tend to appear as the first region in which the homology is
constant, or ``stable'' above the discreteness scale.  Below this
scale, the homology groups fluctuate rapidly as a function of the
thickening parameter.  This provides a necessary though not sufficient
criterion to test for manifoldlikeness of a causal set.

\end{abstract}

\section{Introduction}
\label{introduction}

Any approach to quantum gravity which assumes an underlying Plank
scale spacetime discreteness also requires, alongside, a description
of how continuum topology and geometry arise from the discrete
substructure.  Since the continuum description is apparently robust
down to sub-nuclear scales, the effects of discreteness must be hidden
from standard physics, though the possibility that its signatures can
leak out to scales accessible to current observations and experiments
has not been ruled out.  Replacing spacetime with a collection of
discrete elements cannot suffice, since the only natural topology is
the discrete one, and the only natural geometry is one which makes an
element infinitely far (or close) to every other element.  From purely
continuum considerations one is then led to conclude that the discrete
elements must, at a minimum, be supported by additional relations.

For a theory based on Lorentzian spacetimes, the most natural relation
is the causal relation. The importance of the causal relation to
continuum Lorentzian geometry is emphasized by Malament's result that
the causal structure determines the conformal class for strongly
causal spacetimes \cite{malament}. The conformal factor is the
remaining degree of freedom, and is determined by the local volume
element.  The causal set theory (CST) approach to quantum gravity
gives primacy to the causal relations while assuming a fundamental
discreteness.  The discrete substructure that replaces spacetime is
taken to be a locally finite partially ordered set or causal set,
where the order relation maps to the causal relation in the continuum
approximation of the theory \cite{causets2, valdivia, Henson:2006}.

In CST, the approach to the continuum has a concrete prescription in
terms of a {\sl faithful embedding} $\Phi: \causet \rightarrow
\spacetime$ from a causal set $\causet$ to a spacetime
$\spacetime$. If such a $\Phi$ exists, then $C$ is said to be
approximated by $\spacetime$.  $\Phi$ is an order preserving map which
preserves, on average, the local correspondence between cardinality
and spacetime volume.  By this we mean that the events in $\manifold$
to which the elements of $\causet$ are mapped arise from a Poisson
process at some given density $V_c^{-1}$. Thus, the average number of
causal set elements in a given spacetime volume $V$ is given by
$\langle N \rangle=V/V_c$.  Putting this together with the Malament
result, one obtains, in the continuum approximation of CST, the maxim
``Order + Number $\sim$ Spacetime''.  This construction allows several
kinematical questions to be addressed within CST without explicit
reference to the dynamics\footnote{A dynamical law may be expressed with CST,
  for example, as a (quantum) measure on the space of histories, e.g.\ as
  arising from a process of sequential growth \cite{csg, qmt}, or in terms of
  an action functional on causal sets \cite{bs,2dqg}.}.

A valuable feature of the random lattice associated with the continuum
approximation is that it also implies local Lorentz invariance
\cite{lli}. The randomness however makes the task of extracting
spacetime topology and geometry from the causal set more difficult
than for a regular lattice in which  each element has a fixed finite
valency.  On the other hand, for a causal set that is approximated by
Minkowski spacetime, the elements have infinite valency, a legacy of
local Lorentz invariance.  This means that there is no useful (local)
definition of nearest neighbours for a given element in such a causal
set. However, the prescription for a faithful embedding is itself
concrete enough that substantial progress has been made in extracting
continuum information from the causal set. In particular, one now has
a reasonable understanding of how spacetime dimension \cite{mm},
time-like and spacelike distances \cite{bg,rw}, and a localised Ricci
scalar \cite{bs} can be constructed intrinsically in a causal set.

The construction of spacetime homology has been addressed in detail in
\cite{homology}, and the current work is a numerical follow-up of the
analytical results. Rather than consider the full spacetime topology,
this construction uses a spatial localisation of the causal set to
extract continuum information. In any causal set $C$, a set of
mutually unrelated elements forms an {\sl antichain}, and a complete
set of such elements forms an {\sl inextendible antichain} $\ta$.
This is the analog of a ``fixed time slice'' and can be used to obtain
a frame-dependant localisation. Since the elements of $\ta$ are
unrelated (or spacelike related in the continuum approximation) it has
insufficient intrinsic topological information.  Instead $\ta$ can be
enriched by borrowing information from its embedding in the causal
set. One way to do this is to ``thicken'' it to some $\tta$ by
including elements that are in its neighbourhood. This {\sl thickened
  antichain} $\tta$ is endowed with a richer topological and geometric
structure compared to $\ta$, and the idea is to use it to compare to
the continuum.

In \cite{tas,homology} spacetime volume $n$ was used as a thickening
parameter.  For an $\ta$ which admits a certain ``separation of
scales'' it was shown that there is a range of $n$ for which the nerve
simplicial complex constructed from $\ttav$ is homological to the
continuum with high probability. As discussed in greater detail in
Section \ref{prelims} this is possible so long as there exists an $n$
such that $\ttav$ is ``thick enough'' locally compared to the
discreteness scale, but also thin enough to admit a sensible
localisation. This separation of scales is guaranteed in the continuum
{\sl limit} with the discreteness scale going to zero as the
sprinkling density is made larger. However, given a fixed sprinkling
density, it is not obvious how generic such antichains are. Thus it
becomes pertinent to ask, for a causal set $C$ which is approximated
by a spacetime,  how likely is it that one will pick an inextendible
$\ta$ in $C$ which admits this separation of scales.  The current work
addresses this question using numerical techniques. Using causal sets
obtained from discretisations of 2d and 3d spacetimes, as well as
those generated by other means, we propose a necessary but not
sufficient criterion for manifoldlikeness of a causal set.

We use a Cactus based causal set computational tool
\cite{cactus,toolkit} to construct causal sets, pick out a thickened
antichain and construct its nerve simplicial complex. Subsequently, we
employ the CHomP homology package \cite{chomp} to calculate the
spatial homology groups for a suite of causal sets. The first class of
causal sets arise from discretisations of the following spacetime
regions: (i) the flat 2d and 3d intervals with topologies $I \times I$
and $I \times I \times I$ respectively, (ii) the 2d cylinder spacetime
with topology $S^1 \times I$, (iii) 3d flat spacetimes with topologies
$T^2 \times I$ and $ S^1\times I \times I$, with different spatial
sizes to examine Kaluza-Klein type effects, (iv) the ``split'' 2d
trousers, $I\sqcup I \rightarrow I$ and (v) the 2d expanding FRW
spacetime with topology $S^{1} \times I $. The second class of causal
sets arise from non-manifoldlike considerations: (i) from the
transitive percolation dynamics which belongs to the classical
sequential growth class of dynamics studied in
\cite{csg,rg} and (ii) the tower-of-crowns poset which is a
``crystalline'' causal set obtained via a regular discretisation of
the 2d cylinder spacetime.

Our results demonstrate that, for causal sets that are approximated by
a class of 2d or 3d spacetimes, the continuum homology groups are
stable (constant) over a large range of thicknesses. While some of the
3d computations, being more CPU intensive, give only an indication of
stability, the 2d computations allow us to identity quantitative
correlations between the causal set and the continuum. In the 2d case,
for each causal set obtained from a discretisation of a spacetime from
the above set, we use an ensemble of 100-200 randomly generated
antichains and obtain homology as a function of thickening for
each. We find that the continuum homology has a tendency to appear in
a contiguous range of thicknesses. We characterise the stability of
the homology in terms of a dimensionless parameter, the ``stability
ratio'' $\sn$ (see eqn (\ref{stabilityratio}) and the following
discussion) and define a region to be stable only if $\sn \gtrsim 1$.
We find that the continuum homology does {\it not} appear as the most
stable region. However,   when the width of the stable region is
required to be much larger than the discreteness scale, the continuum
homology appears as the {\it first} stable region for a
high percentage of the trials.
Moreover, the homology changes rapidly
before the first stable region begins, and when comparing across
different trials is uncorrelated for small $n$.

While our 3d results support these conclusions, the quantitative
analysis is limited by computational constraints. In particular, only
minimal antichains are used in all the computations. Such an
antichain, made up of the minimal elements of the causal set, 
is extrinsically as flat as possible, since it stays close to the
flat boundary of the sprinkling region. For this choice, a stable
region with the continuum homology starts to appear, but the
computation cannot always be carried out for large enough thicknesses
to establish whether the region is stable. An interesting class of 
causal sets are those obtained from discretisations of Kaluza Klein
type spacetimes, with topologies $T^2\times I $ and $S^1 \times I
\times I$, where the size of the compactified (internal) dimension
$S^1$ with respect to the non-compact (external) spacetime $S^1 \times
I $ or $I \times I $ is varied.  When the compactification scale is
small, of order of the sprinkling scale, we find that its topology
does not show up in the stable homology which matches only that of the
external spacetime.  Thus, one gets an effective 2d topology as
expected. As the size of the compactified region is increased, an
intermediate stable homology begins to arise which suggests  the
presence of an internal manifold.

Apart from validating the analytic results of \cite{homology} our
results suggest a test for manifoldlikeness in a causal
set.\footnote{See \cite{Joe} for a different test for manifoldlikeness
  for causal sets that are approximated by an interval in 2d Minkowski
  spacetime.} Namely, for a causal set $C$, if in a statistically
large sample of randomly chosen antichains in $C$ (i) the homology
groups for the first stable region obtained from each antichain agree
for most of the trials and (ii) for each of the sampled antichains at
thickness of the order of the discreteness scale the homology changes
rapidly but does {\it not} agree over even a small fraction of the
trials, then $C$ satisfies a necessary but not sufficient condition
for manifoldlikeness.  We demonstrate our test on causal sets
generated via a transitive percolation dynamics 
\cite{csg,rg} as well as for the tower of crowns causal sets. We find
that while causal sets generated by a class of transitive percolation
dynamics qualitatively suggest manifoldlikeness, they do not pass the
quantitative tests.
The tower-of-crowns causal set, which we examine analytically, is an
example of a non-manifoldlike causal set in which there is a
consistent first stable homology region, but for a large ``preferred''
set of antichains it begins immediately, i.e., there is no initial
region of varying homology.

Our results and analysis suggest that the criterion put forth may be
more generally valid.  It would be useful to carry out the computations
in higher dimensions as a check, but as the spacetime dimensions
increase, so does the dimension of an average simplex in the nerve.  While
the Cactus based programs are very efficient in generating the nerves
for a large range of thicknesses for causal sets with $O(10^5)$
elements, 
the homology program CHomP baulks (and reasonably so) at
these very large simplices.

We devote the next section to preliminaries, and describe the
computational set up in Section \ref{comp}.  We present the stable
homology results in Section \ref{results} and the tests for
manifoldlikeness in Section \ref{testman}. In section \ref{conclusions} we
discuss our results and some open questions.

\section{Preliminaries}
\label{prelims}

A {\sl causal set} (or {\sl causet}) $C$ is a finite or countable
collection of elements,  along with a binary order relation $\prec$ which is transitive ($x\prec y$ and $y
\prec z $ $\Rightarrow$ $x \prec z$, irreflexive ($x \nprec x $), and
locally finite ($|\past(x) \cap \fut(y)| < \infty$. Here, 
$\past(x)  \equiv   \{y\in C | y \prec x\}$,  $\fut(x)  \equiv
\{y\in C | x \prec y\} $.

A subset $S$ of a causal set $C$ is implicitly endowed with the causal
relation of $C$ restricted to the subset $S$, and is referred to as a
{\sl sub-causal set} or {\sl subcauset} with $S \subseteq C$. The
future and past of an element $x\in S$ restricted to $S$ are denoted
as $\fut(x,S)$ and $\past(x,S)$, respectively.  A {\sl chain} is a
subcauset which is totally ordered so that every pair of
elements is related, and an {\sl inextendible antichain} $\ta$ is a
subcauset of the causal set $C$ which has no relations among its
elements, and for which every element $x \in C$ is related to at least
one element of $\ta$.
%
A {\sl maximal element} $x \in S \subseteq C$ has
$\fut(x,S)=\emptyset$ and a {\sl minimal element} $y \in S \subseteq
C$ has $\past(y,S)=\emptyset$.

Associated with $\ta$ is its ``volume thickened'' neighbourhood, the subcauset
\begin{equation}
\ttav = \left\{x\in (\fut(A) \cup A) \;| \; |\past(x)
 \setminus \past(A)| \leq  n \right\} \;,   
\label{ta_def}
\end{equation}
where $n$ is any non-negative integer, and $\ttao \equiv A$.%
\footnote{This differs marginally from the definition used in \cite{homology},
 where the counting included $x$.}  
In \cite{homology} $\ttav$ was used to prove a correspondence between
the continuum and the causal set homology and hence will play a
crucial role in our analysis. The parameter $n$ corresponds to a scale
and can be used to compare with continuum expectations.

\begin{figure}[ht]
\centering \resizebox{4in}{!}{\includegraphics{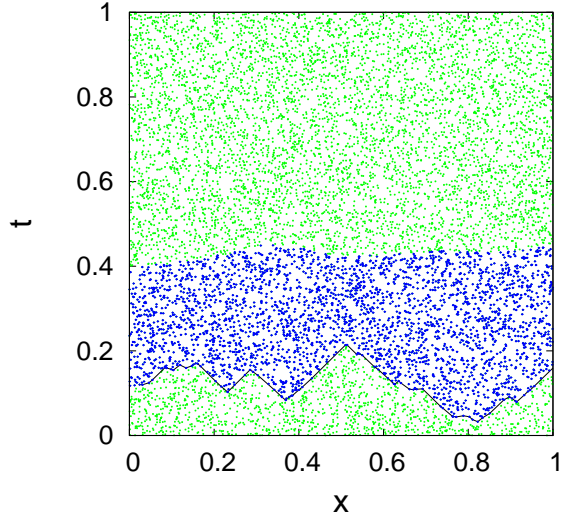}}
\caption{\small A causal set obtained from a sprinkling of $N=20,000$
  elements into a portion of the 2d cylinder spacetime $S^1\times I$,
  with $x=0 \sim x=1$. In order to avoid cluttering, the relations
  have not been drawn, but are determined by the $45 \deg$ light
  cones. A random inextendible
  antichain has non-uniform induced extrinsic curvature (emphasised by
  the jagged line). Thickening it to a volume of $n=1000$ tends to
  uniformise the  curvature.}
\label{smooth}
\end{figure}

Apart from its use in homology studies, this thickening also has the
rather interesting property that it leads to an eventual smoothing out
of the original antichain, i.e., the extrinsic curvature is gradually
``uniformised'' on the antichain obtained from the maximal elements of
the thickening as shown in Fig \ref{smooth}\footnote{In several of the
  figures including this one, we have ``thinned'' the embeddings in
  order to have visual clarity in a black and white printout. However,
  they are best viewed in colour.}. 
The reason for this is that the volume thickening tends to ``fill up''
the valleys, or regions of negative extrinsic curvature regions faster
than it grows the hills, or regions of positive extrinsic
curvature\footnote{This smoothing out of extrinsic curvature appears
  to be at least qualitatively not-unlike a Ricci-type flow although
  it is difficult to construct a local flow equation to make a clear
  comparison \cite{ricci}.}. A past volume thickening has the opposite
effect and uniformises extrinsic curvature to the past.  A qualitative
understanding of this smoothing property in the continuum comes from
comparing the past volumes down to $\Sigma$, of two elements $p, q $
in the future of a spacelike hypersurface $\Sigma$, both a fixed
proper time from $\Sigma$. If $v(p) \equiv I^-(p) \cap \Sigma$
contains a positive extrinsic curvature region and $v(q) \equiv I^-(q)
\cap \Sigma $ a negative extrinsic curvature region (where $I^\pm(x)$
denotes the causal future and past of $x$) then $v(p)>v(q)$.  Thus $q$
will be incorporated into a smaller volume thickening of $\Sigma$ than
$p$, which suggests that the negative extrinsic curvature regions
thicken ``faster'' than the positive curvature regions.

\subsection{Constructing the Nerve Simplicial Complex} 

Starting from a volume thickening $\ttav$ of an antichain, we begin
by constructing ``shadows'' of the pasts of the maximal elements $\{m_i\}$
 of $\ttav$   onto $\ta$     
\begin{equation} 
S^i_n \equiv (m_i \cup \past(m_i)) \cap \ta. 
\end{equation} 
The collection $ \cO_n =\{ S^i_n \} $ covers $\ta$ since $\cup_i S^i_n =
\ta$ which is also locally finite.  The associated nerve simplicial
complex $\cN(\cO_n)$ is constructed by assigning to every element of $\cO_n$
a vertex, and to every $n$-wise intersection, an $n-1$-simplex
\cite{munkres}.

\begin{figure}[ht]
\centering \resizebox{6.0in}{!}{\includegraphics{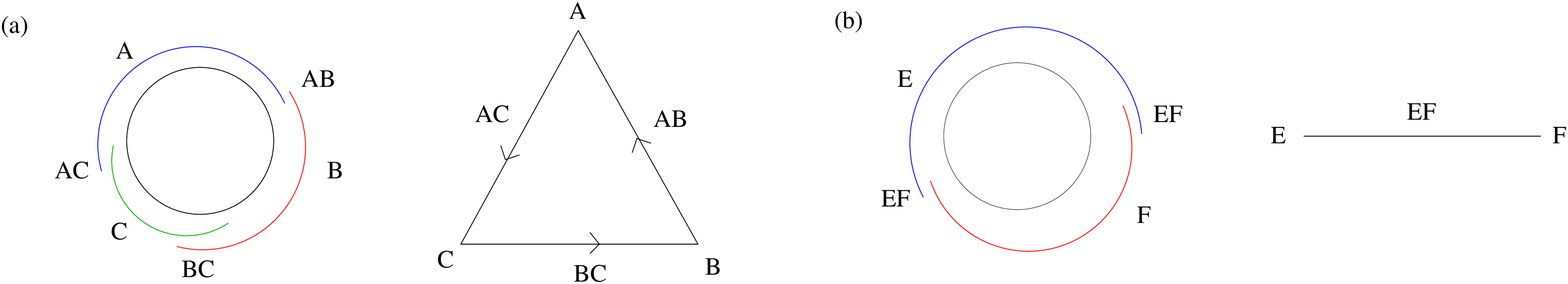}}
\caption{\small Nerves constructed from two different
  coverings of $S^1$. (a) This nerve is homological to the circle 
  since $H_0= \bZ$, $H_1=\bZ$. (b) This nerve is homological to a
  point since $H_0=\bZ$ is the only non-trivial homology group.} 
\label{nerveeg.fig}
\end{figure}

\begin{figure}[ht]
\centering \resizebox{4.0in}{!}{\includegraphics{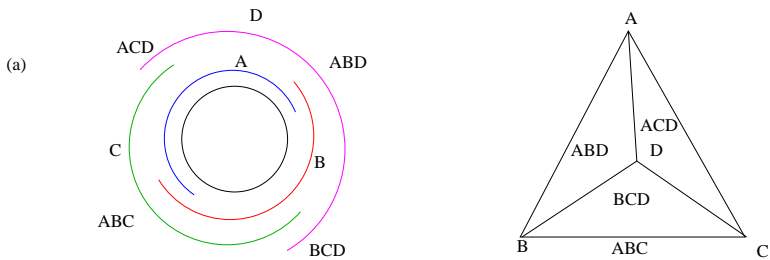}}
\caption{\small A covering of $S^1$ made up of 4 non-convex
  subsets. The three way intersections are all non-zero, resulting in
  $4$ 2-simplexes which make up the boundary of a
  tetrahedron. However, there is no 4-way intersection, so the
  interior of the tetrahedron is left empty. Thus, the non-trivial 
  homology groups are $H_0= \bZ$ and  $H_2=\bZ$, even though the
  covering is of a 1-dimensional manifold.} 
\label{h2.fig}
\end{figure}

Fig \ref{nerveeg.fig} gives an example of two different nerves
constructed from two different locally finite coverings of the circle.
These can be thought to arise from a continuum version of the shadow
construction for a simple spacetime like the flat cylinder $S^1 \times
\re$, where $ds^2=-dt^2 + dx^2$, and $x=0 \sim x=1$.  Starting with
the $t=0$ slice, we can volume thicken to the future.  As shown in
\cite{homology} the future boundary of such a thickening is itself
homeomorphic to $S^1$, and one can pick a few points on this to obtain
a finite shadow cover of the $t=0$ slice. For small thickness, it is
possible to stay within the so called {\sl convexity volume} $v_c$
associated with the slice, which is the largest thickening for which
the shadows are convex subsets of $S^1$.  For such finite coverings, a
theorem due to De Rham and Weil tells us that the nerve simplex is
homotopic to $S^1$ \cite{derham-weil}. However, once the convexity
volume is passed, the simplicial complexes are not necessarily
homotopic to $S^1$. A  useful example of how higher dimensional
homology can arise from such a construction is given in Fig
\ref{h2.fig}.

Past the convexity volume is another important regime associated with
the ``cosmological scale'' $v_\lam$. This is the thickening at which
at least one of the shadows $S_c$ encompasses all of the initial
hypersurface. Such a scale can be infinite as in the case of Minkowski
spacetime, but is finite for compact spacetime regions. Not only
does $S_c$ intersect all other shadows, but also all their
intersections. This means that if the nerve contains a non-trivial
cycle without the vertex $S_c$, its addition must necessarily collapse
it to a trivial cycle. Hence the existence of such an $S_c$ ``washes''
out non-trivial homological information contained in the spatial
slice.  This cosmological scale is therefore the thickening limit
beyond which useful spatial homological information cannot be
extracted.

\begin{figure}[ht]
\centering \resizebox{4.0in}{!}{\includegraphics{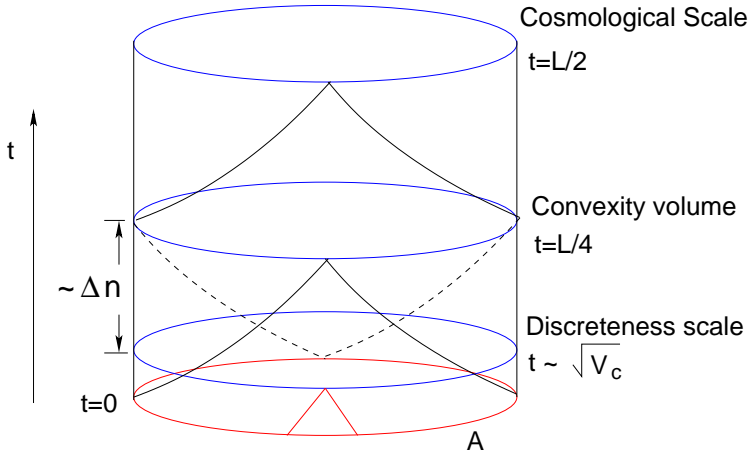}}
\caption{\small The convexity volume and the cosmological scale for a
  $t=0$ slice in the 2d cylinder spacetime with spatial size $L$ are
  shown. For a causal set $C$ which is approximated by this region of
  the spacetime, the minimal inextendible antichain $A$ in $C$ is
  approximated by the $t=0$ slice.  While the region of first stable
  homology for the continuum slice is $t \in (0, L/4)$, that for $A
  \subset C$ lies in the region $t \in (\gtrsim \sqrt{V_c}, L/4)$ where
  $\sqrt{V_c}$ represents, in this case, the discreteness timescale.}
\label{mesoscale.fig}
\end{figure}

In \cite{homology} it was shown that, for a causal set that is
approximated by a globally hyperbolic region of spacetime at a given
sprinkling density $V_c^{-1}$, the nerve $\cN(\cO_n)$ is homological
to the continuum for a large range of $n$'s, as long as $A$ satisfies
a certain separation of scales. In analogy with the continuum, $n$
must be less than the convexity number $n_c \equiv v_c/V_c$.
Moreover, not only should $n \gg 1$, i.e., be far from the
discreteness scale ($n \sim 1$), but also be large enough that $\ttav$
is not too ``thin'' in patches; since the thickening is uneven if $A$
has a non-uniform extrinsic curvature to start with, the maximal
elements of $\ttav$ may continue to lie close to $A$ even for $n \gg
1$. This determines an additional scale $n_0$, related to a minimal
proper distance. The separation of scales requirement is then $1 \ll
n_0 \ll n \ll n_c$, where the $\ll$ are required to take into account
the possibility of large fluctuations. For example, the volume
containing $n_c$ sprinkled elements can be much larger than the
convexity volume $v_c$ beyond which the continuum results are not
valid. The separation of scales therefore allows the possibility of a
large range of $n$ for which $\cN(\cO_n)$ is homological to the
continuum.  Fig  (\ref{mesoscale.fig}) shows how the separation of
scales manifests itself in the 2d flat cylinder spacetime with
topology $S^1 \times I$.  

If it can be shown that inextendible antichains in such a
causal set $C$ {\it generically} satisfy the separation of scales,
then the appearance of a stable homology can conversely be taken to be
an indicator of manifoldlikeness.  The analytical results, while
robust, do not help address this question.

\subsection{The Computations} 
\label{comp} 

Our primary aim in this work is to further the idea that stable
homology can be used to test for manifoldlikeness of a causal
set. Given the complexity of causal sets, such a test requires the use
of numerical tools described below.

The construction of the nerve simplicial complex uses a causal sets
toolkit \cite{toolkit} within the Cactus high performance computing
framework \cite{cactus}.  The toolkit consists of a number of modules
called ``thorns'', which provide various functionalities needed to
perform computations with causal sets.  The first step is to choose a
spacetime $\spacetime$ from which a causal set $C$ can be
obtained. Next, we sprinkle into the spacetime, which yields a number
of causet elements sampled from a Poisson distribution with mean $N$.
(It is to be understood below that phrases such as ``a sprinkling of $N=20,000$
  elements'' means we sprinkle with $N$ set to 20,000, not that the causal
  set necessarily contains 20,000 elements.)
The induced causal relations between the elements are then computed to 
obtain $C$.  This latter process requires a detailed knowledge of the
causal structure of $\spacetime$ which is not always readily
available. Therefore the suite of spacetimes used is currently limited
to a set whose causal structure is explicitly known.  The software
provides for sprinkling into a variety of topologies of conformally
flat spacetime, for any spacetime dimension up to 8+1.  It also
provides many fundamental set operations which are needed for the 
computations described in this paper, such as computing pasts and
futures, unions and intersections, and minimal and maximal elements.

Of course, one also wants to generate causal sets in other ways 
besides sprinkling into spacetimes,
since this avoids an a a priori assumption of manifoldlikeness. The
causal sets toolkit provides several examples of such causal sets,
some of which are generated via so-called sequential growth dynamics
\cite{csg}. 

Once a causal set $C$ has been generated, 
an inextendible antichain $A$ is picked in one of two ways. The
simplest choice is the set of minimal elements. This ties $A$ to the
choice of the arbitrarily defined bounding frame for the sprinkling,
and is therefore unnatural. A more natural and robust procedure is to
choose a randomly generated inextendible antichain, and to
subsequently consider a large number of such choices, to account for
statistical fluctuations.

Since these random antichains are used in a crucial way in the 2d
computations, we discuss it in some detail.  We use the following
algorithm for selecting an inextendible antichain.  First, select an
element from the causal set at random with a uniform distribution, but
restrict this first selected element of the antichain to have label
$\leq {M}/2$, where the labels of the elements are in $[0, \ldots,
  {M-1}]$ ($M$ is the total number of elements in the causal
set). The labels of the elements are given in the order of their time
coordinates for sprinklings, or as described in Section \ref{testman}
for non-sprinkled causal sets.  The labeling is always a {\sl natural
  labeling}, meaning that if $i \prec j$ in the causet order then
their labels satisfy $i<j$.  Thus this restriction has the effect of
causing the antichain to tend to live in the lower half of the causal
set, which is useful because we always thicken it to the future.

\begin{figure}[ht]
\centering \resizebox{4.5in}{!}{\includegraphics{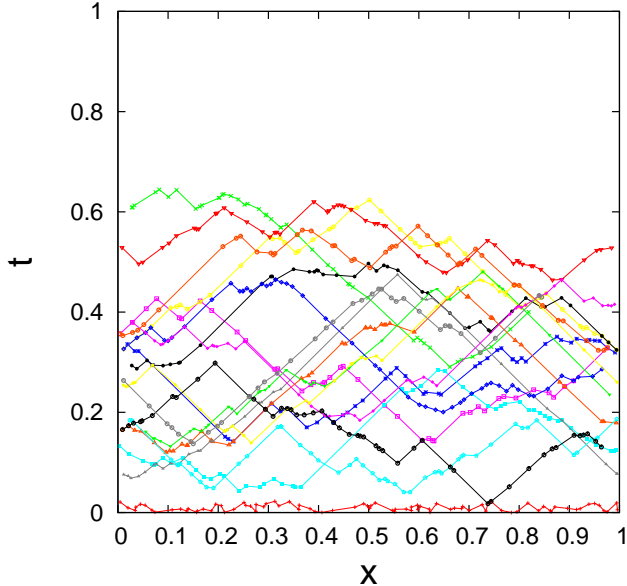}}
\caption{\small A selection of randomly generated inextendible
  antichains in a causal set obtained from a sprinkling of $N=5,000$
  elements into the 2d cylinder spacetime $S^1\times I$, with $x=0
  \sim x=1$. The minimal antichain (not randomly generated) hugs the
  $t=0$ line. We have used lines to connect the elements of the
  individual antichains for visual clarity.}
  \label{sample_antichains}
\end{figure}

Next, select a second element which is unrelated to the first, again
with a uniform distribution on the eligible elements (i.e. the set of
elements unrelated to the first, irrespective of their labels). Then
select a third which is unrelated to the first two, again with a
uniform distribution.  Repeat until there are no elements remaining
which are unrelated to any element selected thus far.  This will
select an inextendible antichain at random, from any causal set.
Fig  \ref{sample_antichains} depicts a collection of such random
antichains. Hugging the $t=0$ surface is the minimal antichain.
The figure indicates that  the algorithm tends to pick out 
antichains with highly varying induced extrinsic curvature, and hence 
those that are ``almost null'' with respect to the preferred slicing
of the spacetime. These antichains seem to be  the least likely to
satisfy the analytical criterion for separation of scales, since
``thin'' patches may persist up to the convexity volume scale. 
In the case of an embedding into the cylinder spacetime $ds^2=-dt^2 +
dx^2$, with $x=0 \sim x=1$ and topology $S^1\times I$, let us consider
the extreme example of a two-element antichain for which the nerve 
construction is inadequate; the only two possible homologies are
$H_0=\Z^2$ or  $H_0=\Z$, both with $H_1=0$.

We use this example to argue that the probability distribution on
inextendible antichains imposed by the above algorithm is far from
uniform, and in fact tends to favor more `pathological' antichains,
which have fewer elements and wider `almost null' segments.  To see
this, for an arbitrary antichain $A$, we define a quantity $n_A(x)$ to
be the number of inextendible antichains which include all of $A$ and
$x$ as well.  For $A$ equal to a single element $x$ in $S^1\times\R$
spacetime, one expects to find a $y$ such that $n_A(y)=1$.  If we
wanted to select each inextendible antichain with equal probability,
then, if $x$ were the first element of our antichain, then we would
have to weight $y$ with a much smaller probability than some other
element $z$ for which $n_A(z) \gg 1$.  However, the above algorithm
does not.  This suggests that the algorithm is weighted toward smaller
inextendible antichains, which contain numerous almost-null segments,
and intuitively are expected to have a larger extrinsic curvature. It
is therefore all the more challenging to test the homology
construction on these antichains.  It is in this sense that we claim that our
results are generic.

Having chosen the antichain, the toolkit contains a module which
constructs the thickened antichain.  As discussed earlier, the
thickening $\ttav$ of a randomly chosen $A$ tends to uniformise the
extrinsic curvature of antichain $A_1'$ constructed from maximal
elements of $\ttav$ for large enough $n$.  The antichain $A_1'$ can
then be completed using the uniform distribution as described above to
obtain an inextendible $A_1$, from which a new set of thicknesses
$T_n(A_1)$ can be constructed. 
The randomness associated with $A$ and its associated non-uniform
extrinsic curvature is thus at least partially tamed. While this
procedure appears promising, it is not a a priori obvious how large $n$
needs to be to uniformise a randomly chosen antichain. In order to
avoid introducing further arbitrary parameters, we make do with the
first thickening $\ttav$.

Next, having picked a large enough thickness $n_m$, the causal sets toolkit
constructs a nerve simplicial complex for each $\ttav$, $n\in [0,
  n_m]$, and stores it in a format that can be accessed by the CHomP
computational homology package \cite{chomp}.  The core CHomP engines
use a variety of cubical homology algorithms to compute homology
groups.  We use the ``homsimpl'' program, which acts as a front end to
CHomP for abstract simplicial complexes.  Before doing so, we simplify
the nerve simplicial complex with a Perl program developed by
Pawel~Pilarczyk which removes redundant vertices. This reduces the
average size of the simplices and thus improves the computational
efficiency substantially. In performing a large number of trials for a
given causal set, finding  their  constant homology regions  and
assessing their stability, we made extensive use of  bash programming. 

In the computations we have to decide a priori what the maximum
thickness $n_m$ should be. It suffices to thicken to the cosmological
scale since we know that the homology is always trivial past this
scale. Thus, $n_m$ must be chosen to be larger than all the possible
cosmological scales. This can be done by generating a large sample of
inextendible antichains in pretrials and finding their cosmological
scales.  In many of the trials the thickening is stopped only at or
very close to the cosmological scale. However, as the sprinkling
density increases, it becomes more computationally intensive to reach
the cosmological scale. On the other hand, the cosmological scale is
only a weak upper bound for the convexity volume. Since we are
interested in establishing the existence of stable regions and showing
that the first stable region is correlated to the continuum homology,
it is then enough to thicken the antichain to some $n_m$ which is
large enough to establish a first stable region. This is indeed what
we do for the high density sprinklings.

The computation of homology groups was by far the bottleneck in all
our computations.  As an illustration, consider an $N=5793$ element
causal set obtained from a sprinkling into the flat $T^2\times I$
spacetime. For a `typical' $\ttav$, with $n=45$, there are $5793^{2/3}
\approx 323$ maximal elements in $T_{45}(A)$. Thus, approximately
$323$ sets cover the $323$ or so elements in $A$.  The nerve
simplicial complex has 187 simplices, with dimensions as large as 19
or more (recall this means simply that there are 20 mutually
overlapping sets in the cover).  Since the number of subsimplices of a
single 19-simplex alone is $2^{20}$, one can see that finding the
homology groups of these complexes becomes a large task for any
computer.

A qualitative picture begins to emerge from examining plots of
homology versus thickness for the several examples considered. First,
we notice that torsion coefficients are trivial in all examples and
hence it suffices to consider the Betti numbers. This is undoubtedly a
curiosity of our simulations; while we expect a constant homology
region which reproduces the continuum homology, there are no known
constraints on the homology outside of this region as long as it
remains within the cosmological scales. It is possible that this is
merely an artifact of our choice of torsion free spacetimes and the
maximum dimensions we can handle.  The plots generically show that as
the thickening is increased from the discreteness scale to the
cosmological scale, that there is an initial period of rapidly
changing homology, followed by at least one region of constant
homology, ending with the cosmological scale at which $H_0=\bZ$ is the
only non-trivial homology (see Fig \ref{cyl.fig} for example.).

In order to make a quantitative statement, we first need to define
stability. Does it, for example, suffice for homology to be constant
over 2 thicknesses, or 20 or 200 for us to deem it stable?  We define the
{\sl stability ratio} to be
\begin{equation} 
\sn \equiv \frac{n_{\mmax} -
  n_{\mmin}}{n_{\mmin}} = \frac{\Delta n}{n_\mmin} \label{stabilityratio}
\end{equation}      
where $n_\mmin, n_\mmax$ are the minimum and maximum thickness,
respectively, for a contiguous region in which the homology is
constant. 

A natural definition of stability is to require that (a)
$\sn \gtrsim \cO(1)$ or $\Delta n \gtrsim \cO(n_\mmin)$, thus ruling
out regions in which the homology is constant only in a relatively
fleeting region $\Delta n \ll n_\mmin$. (For the purpose of the analysis
presented here, we interpret the inequality $y \gtrsim \cO(10^x)$ to
mean $y \geq 5 \times 10^{x-1}$.)  However, this does not exclude
constant homology regions that are too close to the discreteness scale
to reproduce continuum features, since for $n_\mmin \sim \cO(1) $, (a)
is satisfied by a $\Delta n \sim \cO(1)$.  To avoid this, we need in
addition a {\sl mesoscale } $m_s \gg 1$ which is a lower limit below
which continuum features are not expected, so that (b) $\Delta n \gtrsim
\cO(m_s)$ as well. A value $m_s = 100$ is one of the lowest
possible values that satisfies $m_s\gg 1$, and is natural as an order of
magnitude estimate. That this choice is not restrictive is obvious,
since it only sets the lower bound for $\Delta n$. Thus, we will call
a region {\sl stable} if (a) and (b) are satisfied with $m_s=100$, or
more explicitly, that $\Delta n \geq 0.5 \, \,n_\mmin$ and $\Delta n
\geq 50$. 

From our simulations it is clear that the continuum homology does {\it
  not} typically appear as the region with the largest stability
ratio.  It does however appear as a stable region in a very large
fraction of the trials. Moreover, in almost as many trials, it appears
as the {\it first stable region}.  This gives us a concrete hypothesis
for manifoldlikeness, tested over hundreds of examples of causal sets
obtained from 2d as well as 3d discretisations. Conversely, it
provides a reliable method of obtaining the continuum homology if it
exists (see Fig \ref{cyl.fig} for example).

In addition to stability, manifoldlikeness is also characterised by
the existence of a rapidly varying homology for $n\sim \cO(1)$.  The
most significant homology in this region is $H_0$, since at such
thicknesses, connectivity has to be first established. In the 2d
examples, the number of disconnected components $k$, where $H_0=
\bZ^k$, rapidly decreases from large values to a steady small value,
as $n$ increases from $1$ to about $10$. Moreover, when comparing
across trials, one needs to check for initial (roughly up to $n=10$),
rapidly changing homologies that are uncorrelated in different
trials. This can be traced to the randomness of the discretisation
more clearly apparent at small scales.

In the 3d examples, computational constraints prevent us from
performing a similar detailed analysis.  In each case, we note that
regions of relatively large stability corresponding to the continuum
homology indeed do appear, but one is unable in all but a few cases to
continue up to $n_\mmax$ and to assess if this is a stable region. For
the few cases that this is possible, the continuum homology does
appear as the first stable region. Moreover, rather than use random
antichains, we use only minimal antichains. Again, to reduce
computation times, in certain cases we calculate not the homology over
$\Z$, but over $\bZ_2$. Test examples indicate that there are no
compromises because of this, since torsion appears to always be absent
from these specific examples.

A test for manifoldlikeness would therefore proceed as follows: For a
finite cardinality causally convex subcauset $C_0 \subset C$ (for
example, an Alexandrov interval in $C$), construct a random
inextendible antichain $A$ in $C_0$, and check for the existence of
stable homology regions in $\ttav$. Repeat for a statistically large
enough sample. Next, check for correlations among the stable
regions. If over several samplings one consistently gets a homology
that is stable, then this is a good indication of
manifoldlikeness. Next, if for a substantially large fraction, the
homology of the first stable region is the same, then $C_0$ would have
passed the stable homology part of our test.  This, for example,
suffices to demonstrate that although qualitatively the homology of
causal sets generated by a class of transitive percolation dynamics
appears stable and suggestive of manifoldlikeness, a statistical
analysis shows that the case is significantly weaker than for the
other examples studied in this paper.  Specifically, the class of
dynamics that we analyze is restricted by the limited range of
parameters ($p=0.4,0.45, 0.5$) that are computationally accessible to
us. Our trials show that the first stable region does not consistently
give us the same homology for different inextendible antichains. Next,
if the stable homology test is passed, then one needs to check for
rapidly changing homology from $n=0$ to at least $n=10$ which are all
uncorrelated in the different trials.  The absence of such a rapidly
changing region of homology for a large class of preferred antichains
in the ``crystalline'' tower-of-crowns causal set which {\it does}
have a consistent stable homology region means therefore that it is
not manifoldlike.

Before proceeding to the next section we note that the above
construction for homology is not unique, from a poset
perspective. Indeed, posets admit a wide range of topologies
\cite{topologies, sstop}.  A more natural choice is that of {\sl chain
  homology} for a finite sub-causal set $C' \subset C$, which assigns
to every $k$-element chain, a $k-1$-simplex. However, numerical
simulations suggest that this homology, though possibly more
intrinsic, does not capture the topology of the continuum.  
In Fig 5 we present the homology groups which arise from a
sprinkling into a cylinder $S^1\times I$ spacetime, as described in section
3.1.2.  We select an inextendible antichain at random, and give the homology
groups which arise from a sequence of thickened antichains, for (a) the chain
homology, and (b) the nerve.  It is clear that the nerve is much more
effective at capturing the topology of the continuum. 
\begin{figure}[hbtp]
  \vspace{9pt}
\centerline{\hbox{ \hspace{-0.5in} 
    \epsfxsize=3in
    \epsffile{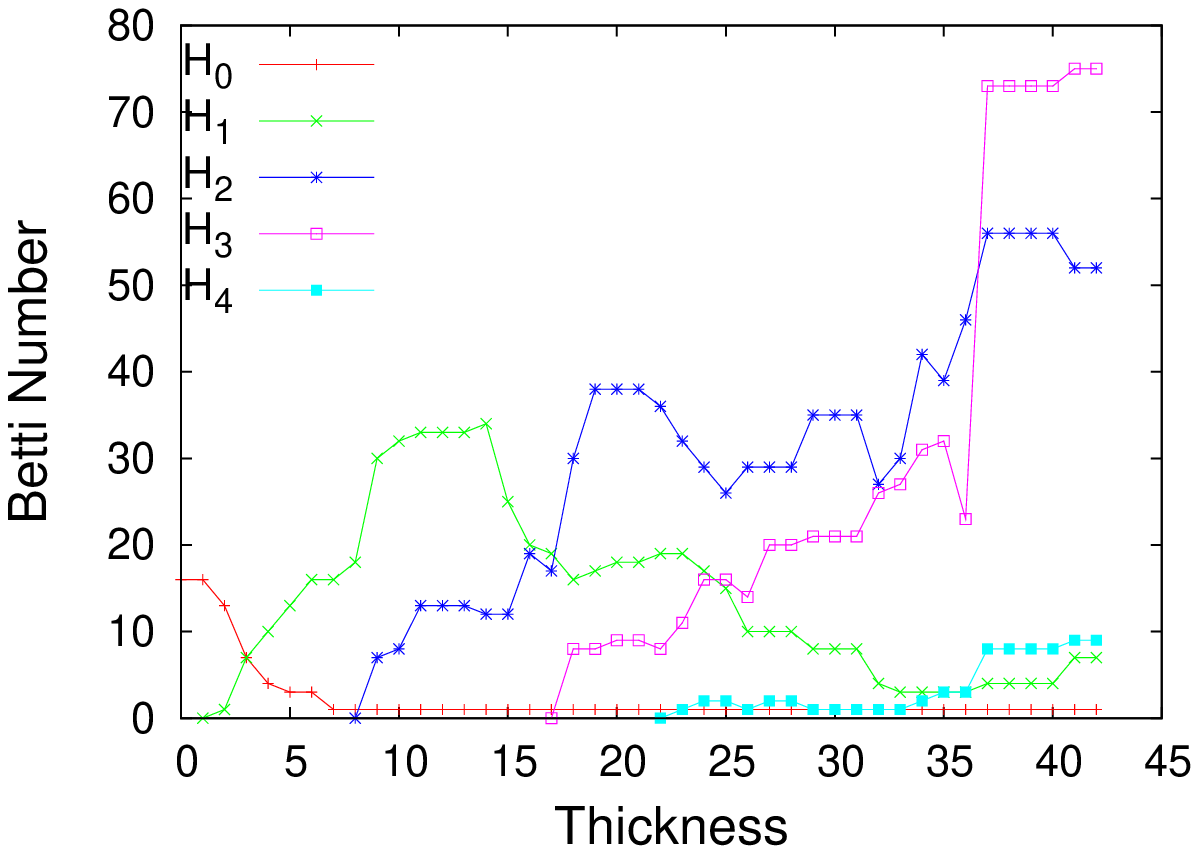} 
    \hspace{0.25in}
    \epsfxsize=3in
    \epsffile{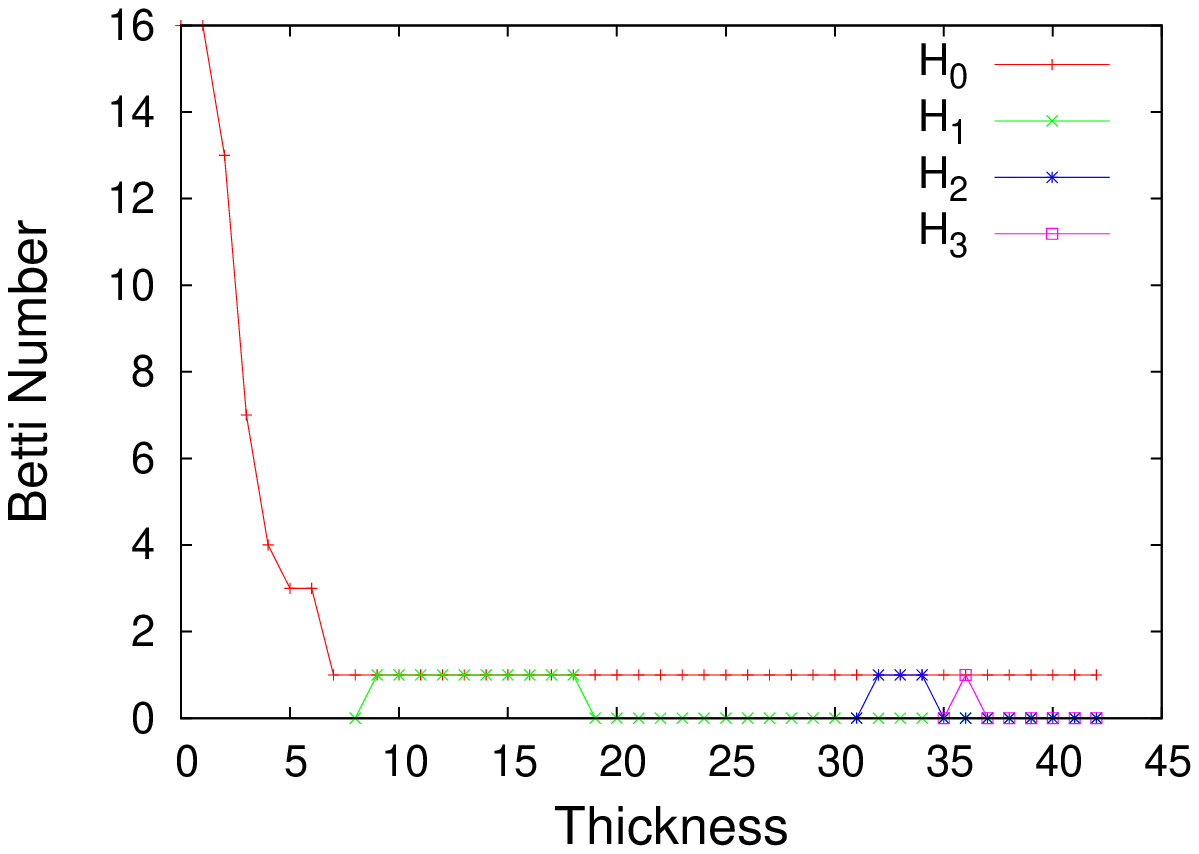}
    }
  }
  \vspace{9pt}
  \hbox
{\hspace{1in} 
{(a)} \hspace{3.1in} {(b)}}  
\caption{ Plots of the Betti numbers as a function of thickness 
    for a random antichain in an $N = 258$ element sprinkling
    into an $S^1\times I$ cylinder spacetime.  (a) shows the Betti
    numbers obtained from the chain homology of
    the thickened antichains, while (b) shows the Betti numbers
    arising from the nerve of the cover generated by the shadows cast
    by the maximal elements.}   
\label{chain} 
\end{figure}

\section{Results in Stable Homology}
\label{results}

In this section we present our main results, using the computational
tools described above, for causal sets that are  approximated by a class of 2d
and 3d spacetimes.  Before doing so, it is useful to first examine the
choice of mesoscale $m_s=100$ more critically.  Our subsequent
analysis depends crucially on this choice,  which in turn suggests that
the first stable region is typically that of the continuum.  

It might seem plausible that as the sprinkling density $\rho$ is
increased, ``spurious'' regions of stable homology appear as the first
stable region suggesting that $m_s$ needs to be modified. Would different
choices of mesoscale give us different first stable homologies? The
answer is yes, when we compare $m_s = 10$ with $m_s= 100$, since
constant homology regions that are stable on scales of order the
discreteness are spurious and do not reflect properties of the
continuum. On the other hand, if we compare $m_s =100$ with an $m_s'
\gg 100$, then differences in the the analysis should arise only if
the convexity volume (for the associated antichain) is of order
$m_s'$. If this is the case, then one should expect no stable regions
with the choice of mesoscale $m_s'$. If $m_s'$ remains sufficiently
smaller than the convexity volume, then the first region of stable
homology will be the same as that obtained using $m_s$.  The reason is
that in our analytical understanding, there are only two scales
between which the stable continuum homology is firmly wedged, namely
the discreteness scale and the convexity scale\footnote{We can ignore
  the additional intermediate scale $n_0$ described at the end of
  Section 2.1, since our focus is on generic antichains. }.  It is
therefore highly unlikely that a spurious first stable region arises
before the continuum homology region. And for the same reason, if one
requires $\Delta n $ to be larger than the convexity volume, then the
first stable region will typically not correspond to the continuum
topology. In Fig \ref{mesoscale.fig} these scales are shown for an
antichain approximated by a $t=0$ slice in the 2D cylinder spacetime,
with $x=0 \sim x=L$.


On the other hand a mesoscale could be determined from other
expectations of the theory, like the existence of a non-locality scale
\cite{nonlocal}. In particular, if a fine-grained causal set $C$ has a
continuum approximation only beyond a certain
coarse-graining\footnote{A causal set can be coarse grained by a
  process of random decimation, in which one removes some fraction of
  the causet elements uniformly at random.} then $m_s' \gg 100 $
within $C$ and one should not expect continuum homology to be
reproduced at smaller scales. Such causal sets will then require tests
for different mesoscales. Equivalently, if $C$ is sufficiently coarse
grained to some $C' \subset C$ for which the continuum approximation
should be valid, then $C'$ should pass our test for manifoldlikeness
with mesoscale $m_s = 100$.  In the causal sets we examine which are
obtained by discretisations of 2 and 3 dimensional spacetimes,
therefore, the lower bound for the mesoscale $m_s = 100$ suffices.

The role of the mesoscale can also be quantitatively understood by
observing how $\sn$ varies as a function of $\rho=V_c^{-1}$ for causal
sets obtained via sprinklings into a particular spacetime at different
densities.  We consider sprinkling densities of $N=1000$ to $N=8500$
in increments of $500$ elements onto the unit 2d cylinder spacetime
\begin{equation}
ds^2= -dt^2 + dx^2 \quad t\in [0,1], x \in [0,1], x=0 \sim x=1. 
\end{equation} 
For each of the causal sets thus obtained, we use the minimal
inextendible antichain. While these antichains are not strictly
coarse-grainings of each other, they are sufficiently close for our
purpose. In each case, the simulations generate the continuum homology
as the first stable region using $m_s = 100$, which shows that this
criterion is indeed independent of $\rho$, as long as the causal set
remains manifoldlike under coarse grainings. 

\begin{figure}[hbtp]
  \vspace{9pt}
\centerline{\hbox{ \hspace{-0.5in} 
    \epsfxsize=3in
    \epsffile{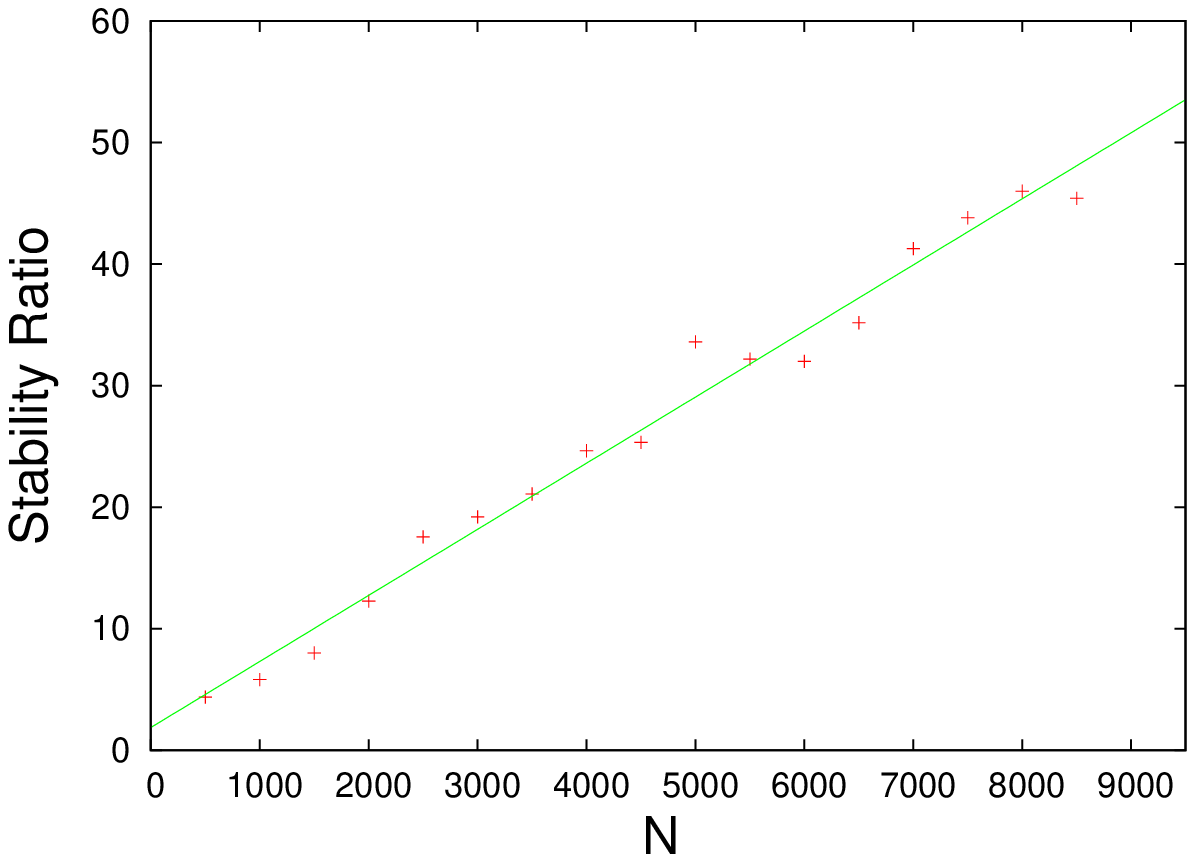} 
    \hspace{0.25in}
    \epsfxsize=3in
    \epsffile{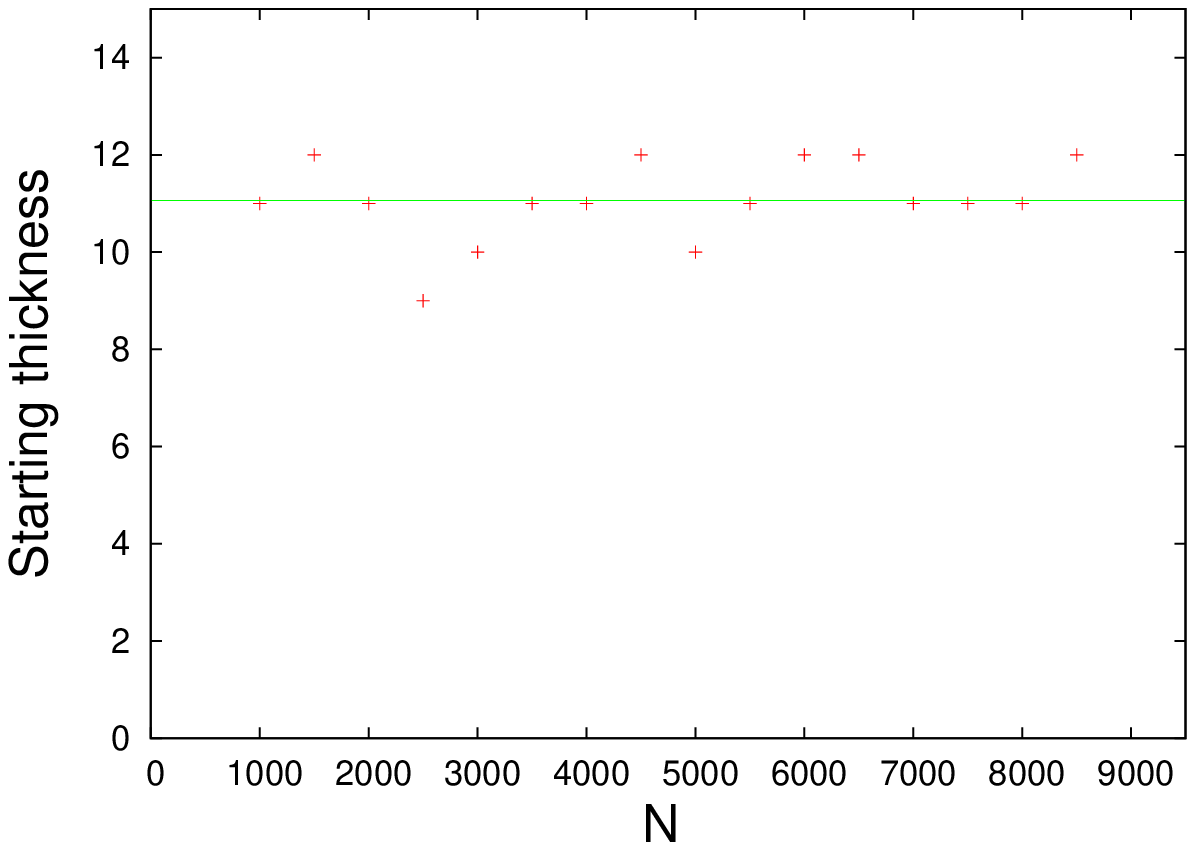}
    }
  }
  \vspace{9pt}
  \hbox
{\hspace{1in} 
{(a)} \hspace{3.1in} {(b)}}  
\caption{(a) A plot of the the stability ratio $\sn$ versus the
  inverse coarse-graining scale $N$ for the 2d cylinder spacetime.  (b) A
  plot of  the starting thickness of the first stable region
  $n_i$ versus the inverse coarse-graining scale $N$.}\label{NvsSR.fig}
\end{figure}

We moreover find a linear relation between $\sn$ and $N$ as shown in
Fig \ref{NvsSR.fig}(a).  This relation can be understood by rewriting
$\sn = \frac{n_f -n_i}{n_i}= \frac{v_f -v_i}{v_i}$, where the $i$ and
$f$ subscripts refer to the initial and final thickening and $v_i= n_i
\rho^{-1}$ and $v_f= n_f \rho^{-1}$ are the associated spacetime
volumes. $v_i$ itself is a function of $\rho$, since as $\rho
\rightarrow \infty$, $v_i \rightarrow 0$. If we take $v_i=\rho^{-1} +
\cO(\rho^{-2})$, then to leading order, $n_i$ is a constant, as seen
in Fig \ref{NvsSR.fig}(b).  On the other hand, $v_f$ being
``macroscopic'' is independent of $\rho$, and hence, $\sn= \frac{\rho
  v_f -n_i}{n_i} = a \rho +b + \cO(\rho^{-1})$, where $a$ and $b$ are
constants. The approximation $\sn \approx a \rho +b$ yields a
reasonable linear fit in Fig  \ref{NvsSR.fig}(a). 

\subsection{2d Spacetimes}

\subsubsection{ Interval Spacetime: $I \times I $ with metric   $ds^2 =
  -dt^2 + dx^2$, $t \in [0,T]$, $x \in [0,x_1]$.}  This is a patch of
Minkowski spacetime, and the computations involved here can be seen as
a ``null test'' for topology. Namely, for any $\spacetime$, if the
spacetime region is chosen to be small enough to be topologically
(though not necessarily geometrically) trivial, then the appearance of a
stable homology should correspond closely to this example. Far from
being a trivial example, it characterises the topological structure of
all spacetimes in the small. While the inextendible antichain
obtained from a sprinkling into this spacetime region cannot be
inextendible in the full space, it is nevertheless sufficient to
construct the requisite localised homology. The only non-vanishing
continuum homology for this spacetime is $H_0=\Z$.

We perform two different classes of computations. In the first case
(a) we take $N=5000$, $x_1=1, T=1$ and thicken each randomly generated
antichain up to $n_m=999$, or the cosmological scale $n_\lambda$,
whichever comes first. In most of the trials, $n_\lambda< 999$, so we
indeed almost always thicken all the way to the cosmological scale. In
the second case (b) we consider a much larger $N$ of $30,000$, with
$x_1=5, T=2$. This gives us a locally dense sprinkling which makes it
computationally expensive to obtain the homology groups up to the
cosmological scales. However, since we are interested in establishing
a correlation between the first stable region and the continuum
homology, it suffices to stop the computation at some reasonably large
thickness $n_m=499$.  For each case, we repeat the calculations with
200 different random antichains on the same sprinkled causal set.
Those antichains which when thickened hit the maximal elements of the
causal set are thrown away in the final analysis, since we imagine physical
causal sets continuing past the maximal elements of our finite simulations. 
\footnote{Had we chosen to thicken to the past (and imagined our
  causal sets to continue past the minimal elements in all cases), we
  would have to do the same for thickenings that hit the minimal
  elements.} This reduces the overall set of useful trials. We will
term a trial {\sl legitimate} if the thickened antichain does not hit
a maximal element before reaching $n_m$ or the cosmic scale.  In
different sets of trials for the different causal sets, we have seen
the number of such legitimate trials fluctuate from $45$ to $99$ in
$100$. In order to achieve a minimum of about $100$ legitimate trials
we therefore find it necessary to generate $200$ random antichains.
We have $124$ legitimate trials for (a) and $130$ for (b).

\begin{figure}[hbtp]
  \vspace{9pt}
\centerline{\hbox{ \hspace{-0.5in} 
    \epsfxsize=4.5in
    \epsffile{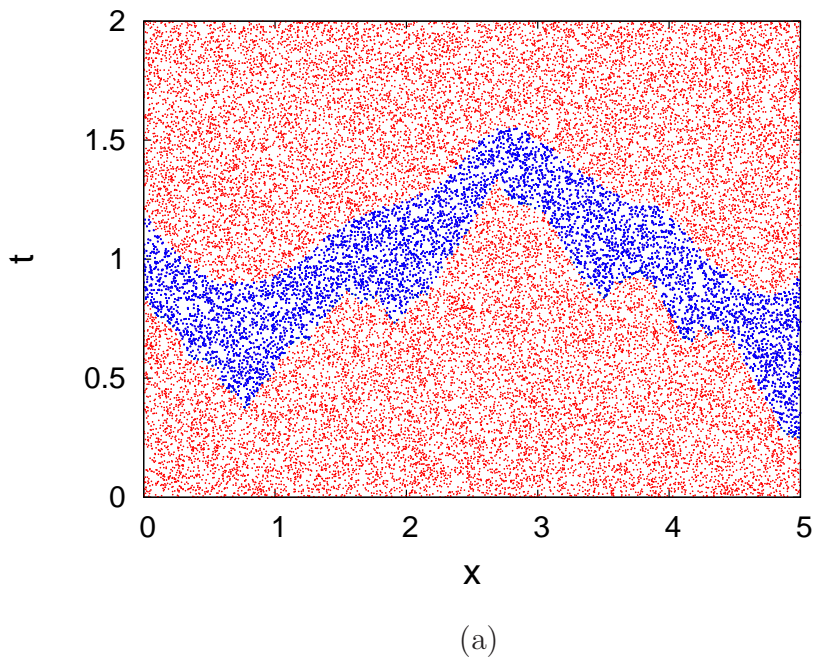}
        }
   }\vspace{9pt}\hbox
{\hspace{2.9in}{(a)}}
  \vspace{9pt}
\centerline{\hbox{ \hspace{-0.5in} 
    \epsfxsize=4.5in
    \epsffile{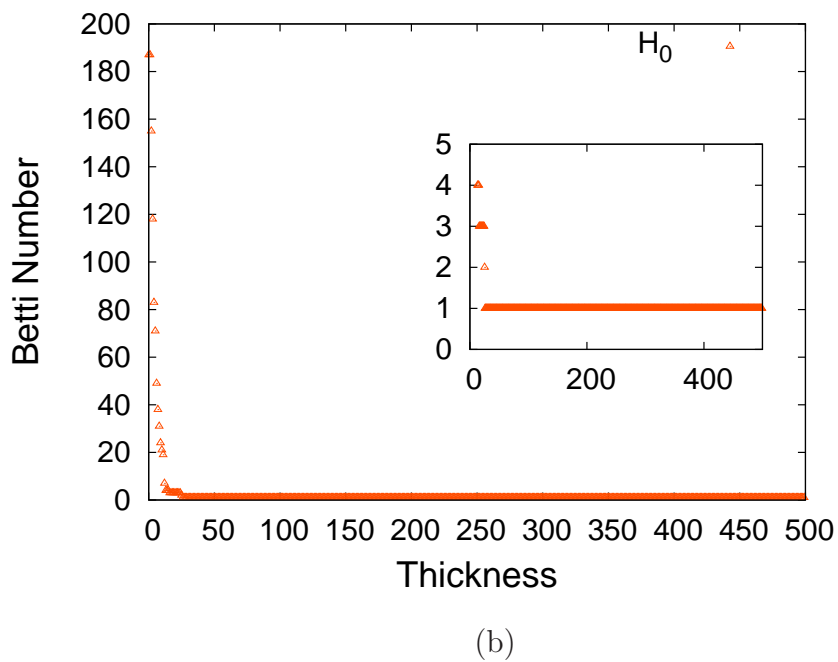}
    }
  }\vspace{9pt}
  \hbox
{\hspace{2.9in} 
{(b)}}  
\caption{ (a) An $N=30,000$ element causal set sprinkled into a ``squat''
  2d interval spacetime, with $x\in [0, 5]$, $t \in [0,2]$. A randomly
  chosen antichain is thickened up to $n=500$. The thickening stops
  well below the cosmological scale. (b) A plot of the betti numbers
  versus thickness. The continuum homology $H_0=\bZ$ appears as the
  first (and only) stable region. All other homology groups
  vanish.}  \label{2dmink.fig}
\end{figure}

In order to demonstrate the importance of the mesoscale $m_s =100$ we
first perform the stability analysis without the additional
restriction that $\Delta n \gtrsim \cO(m_s=100)$. In (a) the continuum
homology appears as the first stable region in $65$ out of the $124$
legitimate trials in (b) in $75 $ of the $130$ legitimate trials.  We
regard this as too poor a result. Indeed, a close examination shows
that one is counting spurious stable regions which may be stable for a
minimum of only two consecutive thicknesses!  On the other hand, using
the mesoscale of $m_s = 100$ gives us an agreement with the continuum
of $100 \%$ for (a) and $127$ out of $130$ for (b). In all the trials,
$H_0$ varies rapidly from $n=0$ to $n=10$ and the behaviour is
distinct for each of the trials. Also, notably, in both cases, the
only non-vanishing homology group is $H_0$. Thus, there is only one
stable region all the way up to the maximal thickness. In the set of
trials (a) this is also true for the cases in which $n_\lambda <999$
and hence the cosmological scale is reached.  We show figures from a
sample trial from set (a) in Fig \ref{2dmink.fig}.

\subsubsection{Cylinder Spacetime: $S^1 \times I $ with metric   $ds^2
  = -dt^2 + dx^2$, $t \in [0,T]$, $x=x_0 \sim x=x_1$.}  This is a
section of the cylinder spacetime, foliated by spatial $S^1$'s. Again,
as in the previous case, we consider two different sets of trials. In
(a) $N=5000$ with $x_0=0, x_1=1, T=1$, and the thickening is taken
up to $n_m=999$ or the cosmological scale $n_\lambda$, whichever comes
first, while in (b) $N=30,000$ with $x_0=0, x_1=5, T=2$ and the
thickening stops well before the cosmological scale at $n_m=499$. For
(a), after generating the first $100$ trials we found $99$ of these to
be legitimate and hence stopped the trials, while for (b) $200$ trials
yielded $155$ legitimate ones. Moreover, in (a) the cosmological scale
is reached for all but four of the trials, and hence one obtains a
fairly global characterisation of the homology groups.

\begin{figure}[hbtp]  
  \vspace{5pt}
\centerline{\hbox{ \hspace{-0.5in} 
    \epsfxsize=3in
    \epsffile{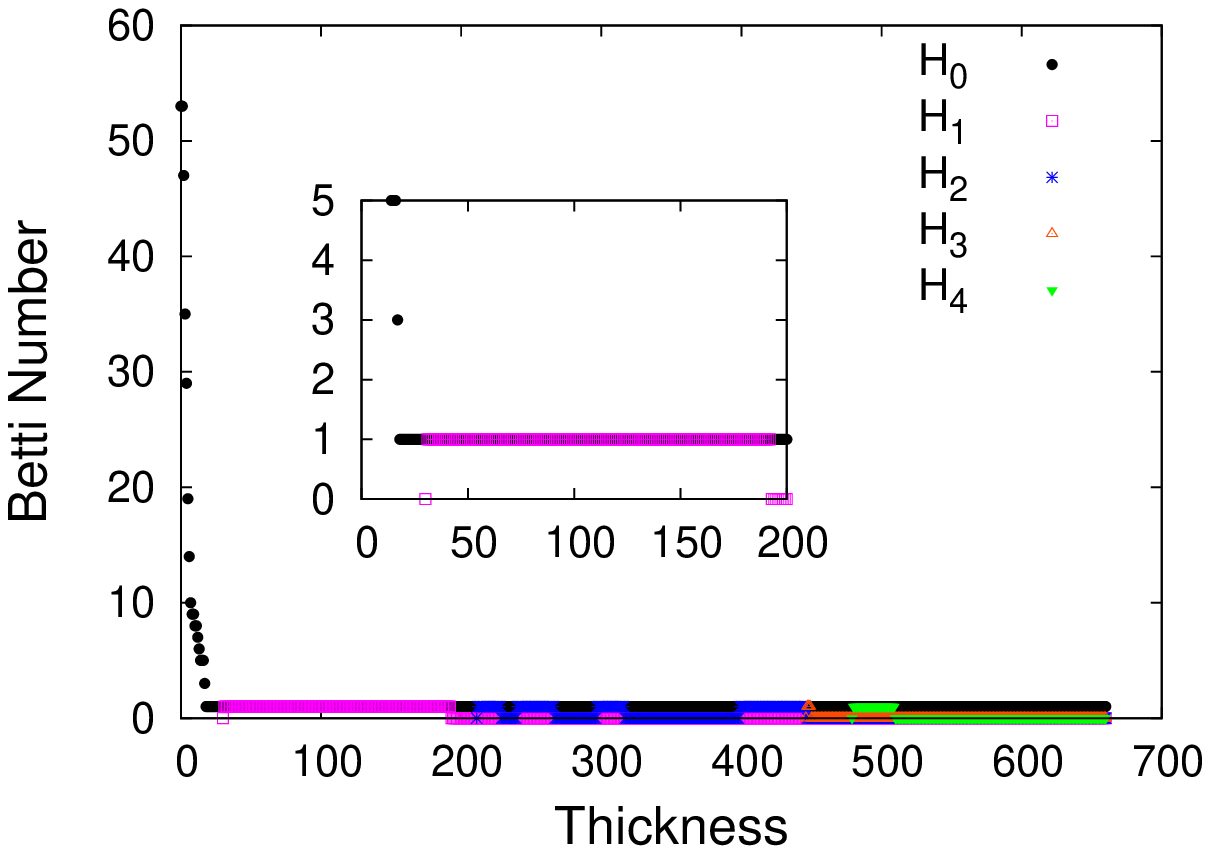}
    \hspace{0in} 
    \epsfxsize=3in
    \epsffile{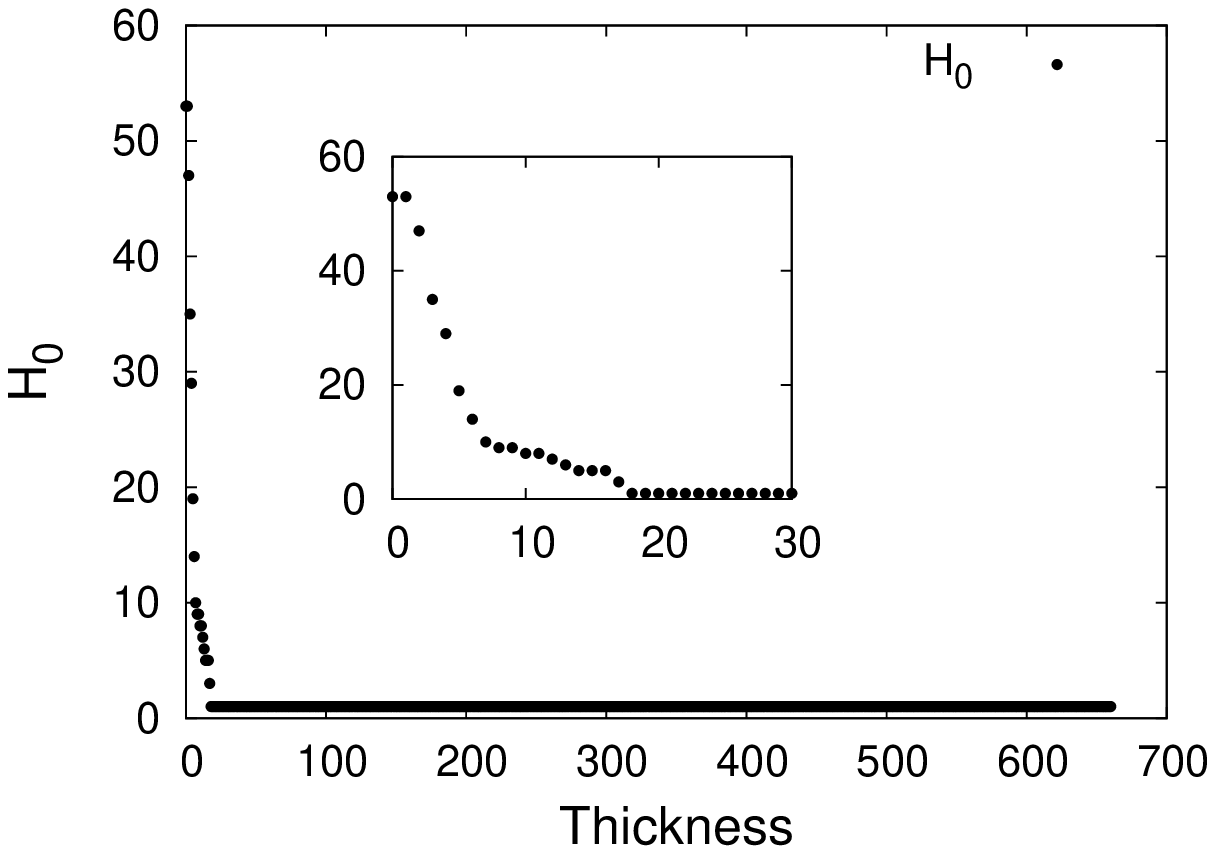}
    }
  }
  \vspace{5pt}
  \hbox
{\hspace{1in} 
{(a)} \hspace{3in} {(b)}}
 \vspace{5pt}
\centerline{\hbox{ \hspace{-0.5in} 
    \epsfxsize=3in
    \epsffile{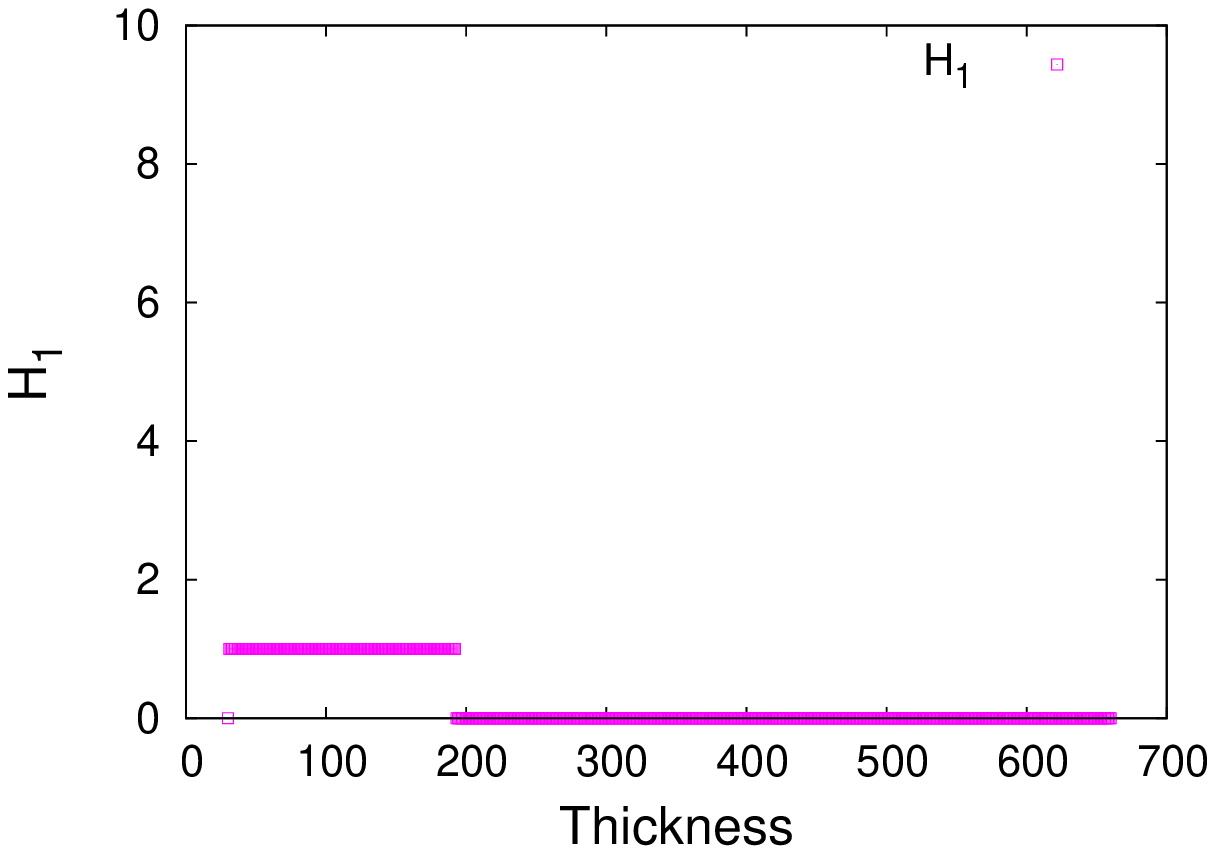}
    \hspace{0in}
    \epsfxsize=3in
    \epsffile{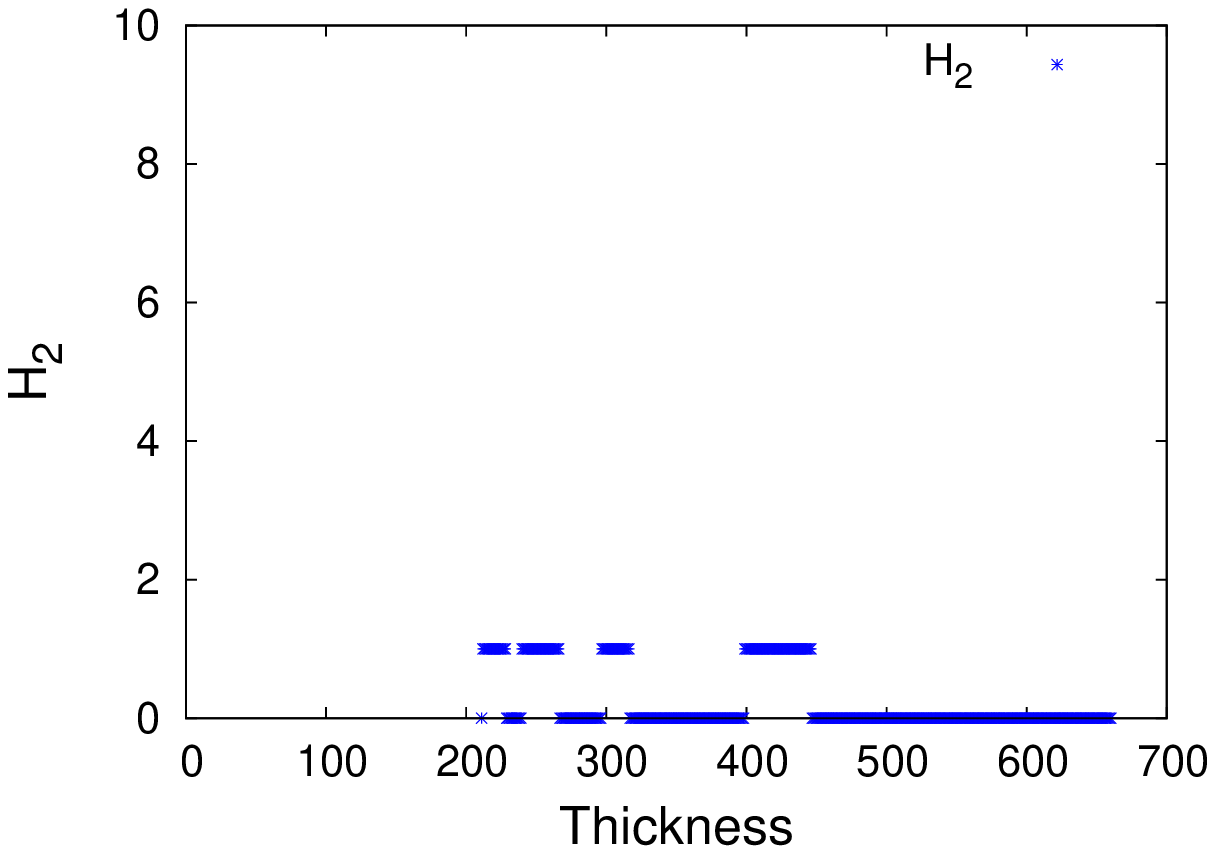}
    }
  }
  \vspace{5pt}
  \hbox
{\hspace{1in} 
{(c)} \hspace{3in} {(d)}}
\vspace{5pt}
\centerline{\hbox{ \hspace{-0.5in} 
    \epsfxsize=3in
    \epsffile{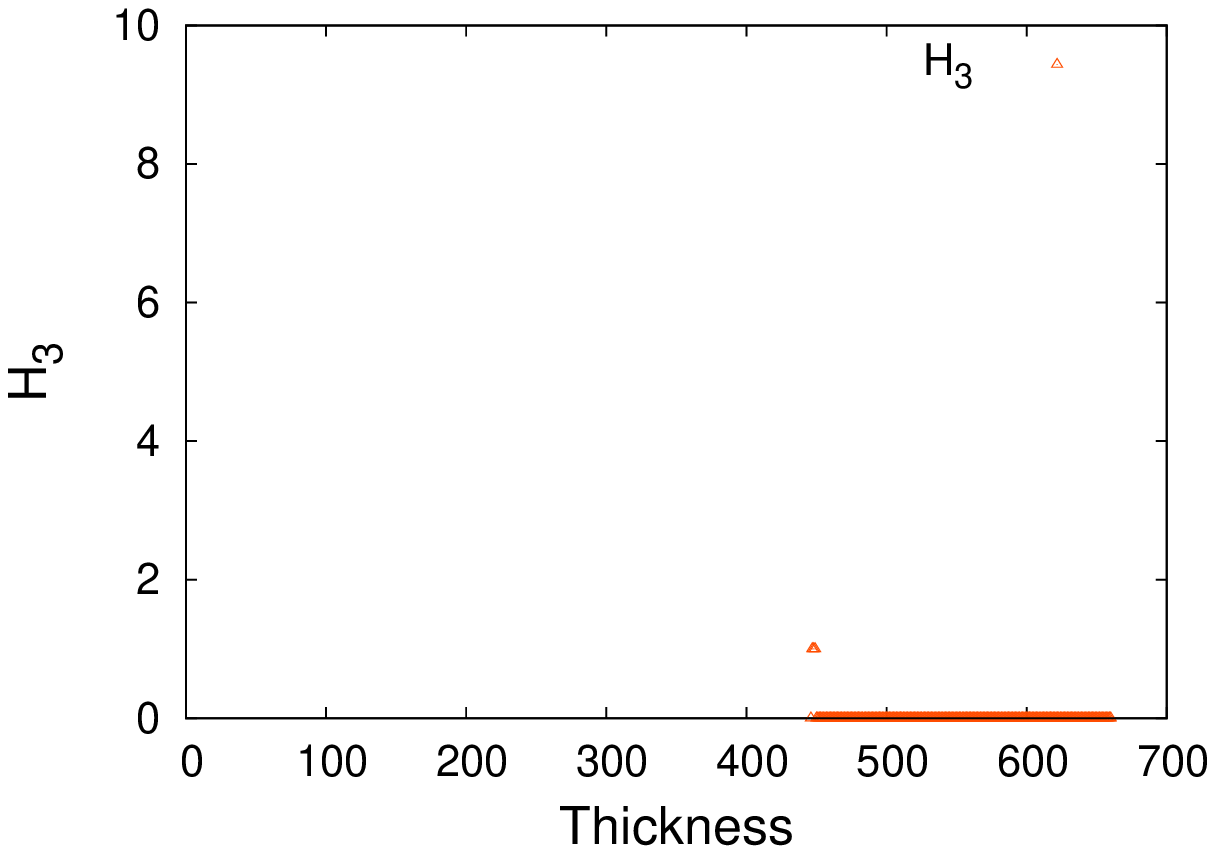}
    \hspace{0in}
    \epsfxsize=3in
    \epsffile{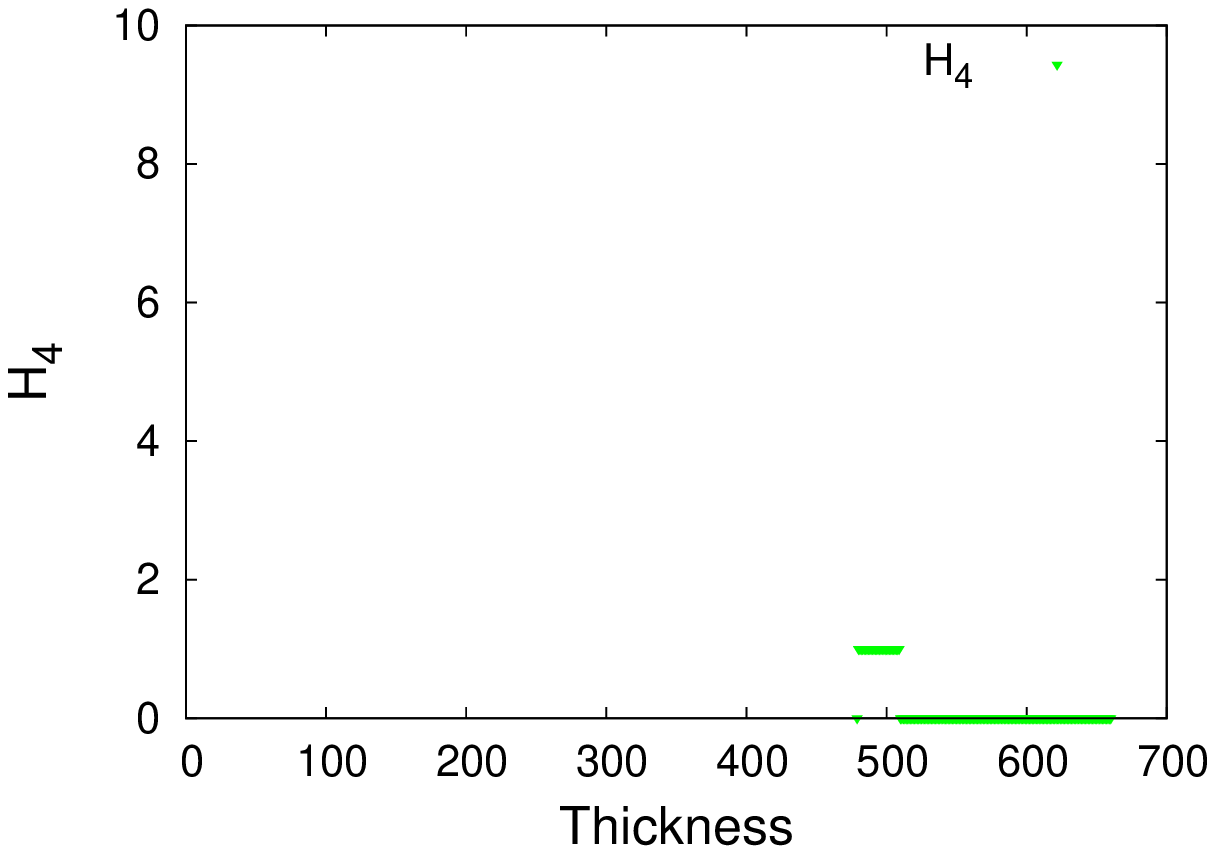}
    }
  }
  \vspace{5pt}
  \hbox
{\hspace{1in} 
{(e)} \hspace{3in} {(f)}}
\caption{Homology from a random antichain in an $N=5000$ element causal set
  sprinkled into a (unit) 2d cylinder spacetime. In (a) all the Betti numbers
  are plotted together as functions of thickness. (b)-(f) resolve this
  graph. } \label{cyl.fig}
\end{figure}

\begin{figure}[hbtp]
\vspace{9pt}
\centerline{\hbox{ \hspace{-0.5in} 
    \epsfxsize=4.5in
    \epsffile{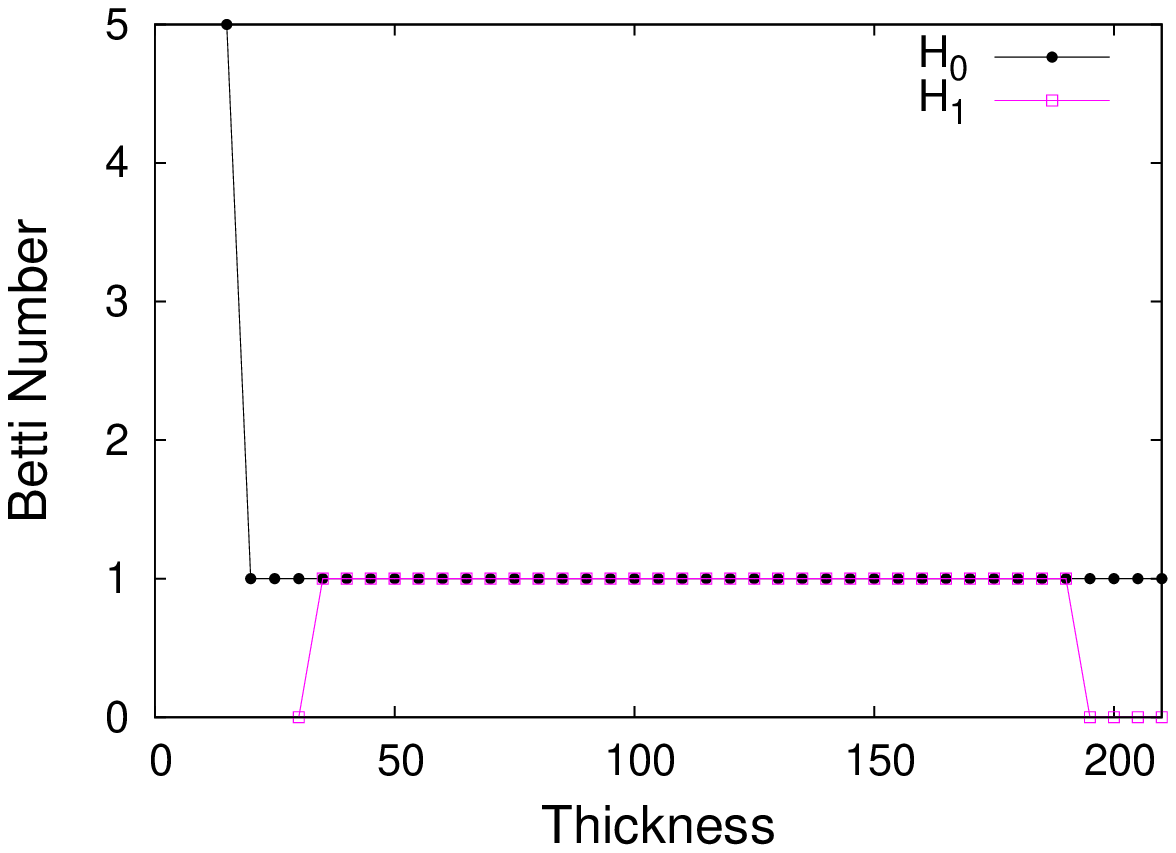}}}
 \vspace{9pt}\hbox
{\hspace{2.75in} 
{(g)}}
\centerline{\hbox{ \hspace{-0.5in} 
    \epsfxsize=4.5in
    \epsffile{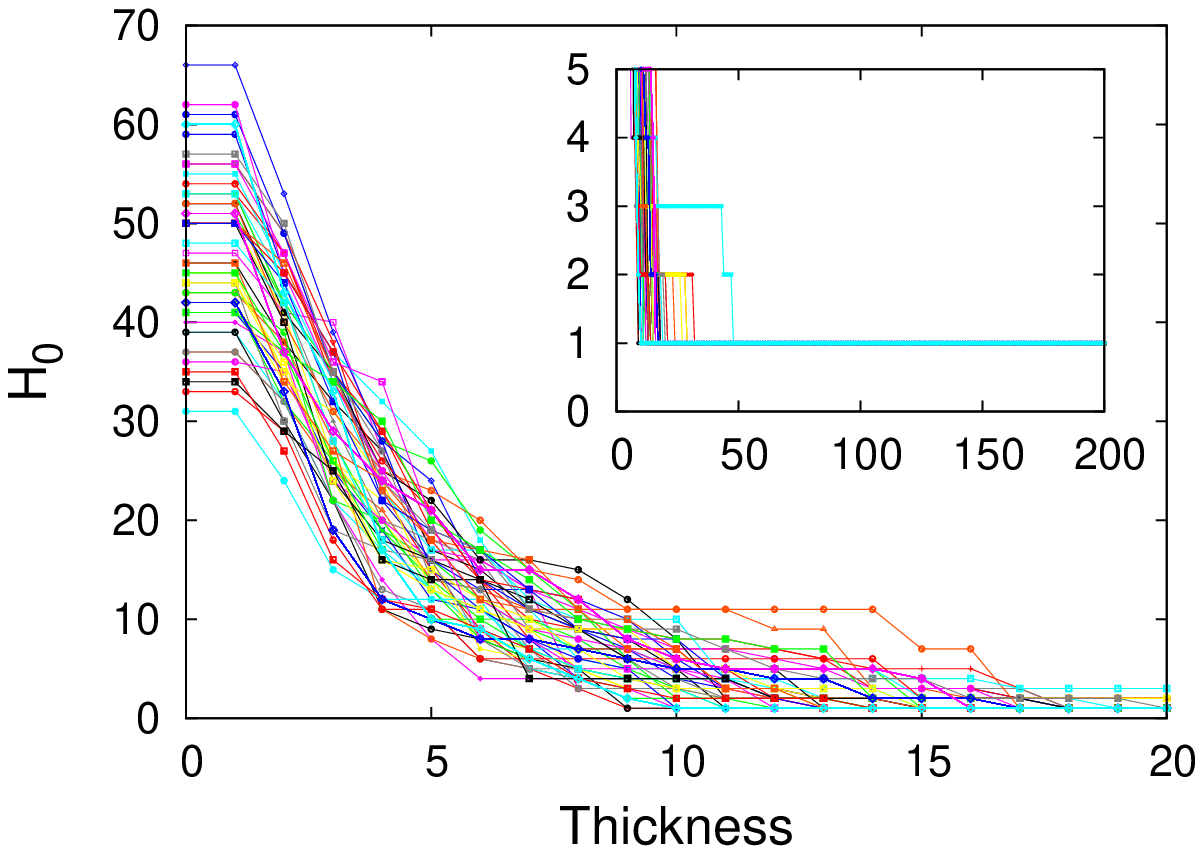}
    }
  }
  \vspace{9pt}
  \hbox
{\hspace{2.75in}  {(h)}}
\caption{(g) A closer look at the stable homology region 
  of Fig \ref{cyl.fig} from $n=12$
  to $n=133$,  with $H_0=\Z$ and $H_1=\bZ$. (h) A  plot showing the fluctuations of $H_0=\bZ^k$ for  $n \in [0,
    15]$ for 50 of the trials. In all cases, while $k$ decreases
  rapidly, the detailed behaviour is distinct for each of the trials.} 
 \label{cyl-zoom.fig}
\end{figure}

Without assuming any mesoscale, the continuum homology appears as the
first stable region in only $47$ of the $99$ legitimate trials for (a)
and $89$ of the $155$ legitimate trials for (b). Assuming an $ m_s =
100$ gives us a vastly improved result of $98$ out of $99$ for (a) and
$154$ out of $155$ for (b). In both sets, for $n=0$ to $n=10$, the
first homology group rapidly varies, and this variation is distinct in
{\it all} the trials.  In case (a) for all the $99$ trials one sees
that several higher homology groups, i.e. $H_i$, $i>2$, can become
non-trivial once the first stable region is passed, but these differ
from one trial to the next.  
In case (b) on the other hand, for all the $155$ trials, only $H_0$
and $H_1$ are non-trivial. This can be attributed to the fact that the
thickening stops before the end of the first stable region. Figures
\ref{cyl.fig} and \ref{cyl-zoom.fig} show the homology groups as a
function of thickness for 
one of the trials.

\subsubsection{ Expanding FRW Spacetime: $S^1 \times I$ with metric  $ds^2=
  t^2(-dt^2 + dx^2)$, $t \in [T_1,T_2]$, $x=0 \sim x=1$.} Although the
topology is that of the cylinder spacetime, it is an important example
of a spacetime with curvature. We again perform 100
computations for $N=15,000$, and assume a mesoscale $m_s = 100$. The
trials run up to $n_m=499$ or the cosmological scale $n_\lambda$,
whichever comes first. We consider two sprinklings, (a) one for
$T_1=0, T_2=5$, which therefore includes the initial singularity, and
(b) another for $T_1=4, T_2=6$. We find that all the trials are
legitimate in both cases. 

\begin{figure}[ht]
\centering \resizebox{5in}{!}{\includegraphics{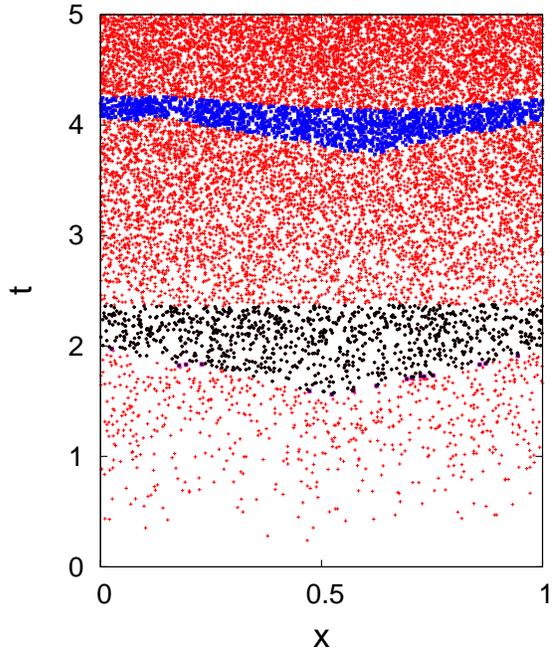}}
\caption{ An $N=15,000$ element causal set obtained from sprinkling
  into a 2d FRW spacetime. Two different antichains $A_1$ and $A_2$
  are used, and both are thickened up to their respective cosmological
  scales $n_\lambda$. Note that that for clarity, the spatial
  direction has been scaled up in comparison with the temporal
  direction. Thus, the light cones are widened. The thickened
  antichain $T(A_1)$ lies closer to the $t=0$ singularity and has a
  small $n_\lambda=75$. The continuum homology does not appear as a
  stable region for this trial. On the other hand the thickened
  antichain $T(A_2)$ is further from the singularity and has a much
  larger $n_\lambda=462$. It has a region of stable homology
  corresponding to the continuum.}\label{badfrw}
\end{figure}

For (a) we find that the continuum homology appears as the first
stable region for only $82$ of the $100$ trials. And for (b) this
improves to $96$ of the $100$ trials. The lower agreement for (a) may
be attributed to the existence of random antichains that lie too close
to the $T_1=0$ singularity, thus preventing it from being
manifoldlike. An example from the trials in (a) which does not
reproduce the continuum homology is shown in Fig \ref{badfrw} as the
lower antichain $A_1$, which exhibits only the trivial homology.  Being
close to the initial singularity, it has a very small cosmological
scale of $n_\lambda=75$ which gives it no time to develop a stable
spatial homology. Another example from the trials (a) which does
reproduce the continuum homology is shown as the upper antichain $A_2$
in the same figure which, being sufficiently far away from the origin,
has a larger $n_\lambda$ of $462$.  In both sets of trials, for all
trials, there is a rapidly varying region of $H_0$ from about $n=0$ to
$n=10$, and the variation is distinct for all the trials. In case (a)
$98$ of the $100$ trials had $n_\lambda<499$, so that the cosmological
scale was reached. As in the cylinder spacetime, some higher homology
groups (up to $H_7$) start to become non-trivial once the first stable
region is passed. In case (b) only $48$ of the trials reached the
cosmological scale, and for $13$ of these trials, one did not reach
the end of the first stable region. For these, all the higher homology
groups, $H_i$ with $i \geq 2$ are trivial.

\subsubsection{The Split Trousers Topology: $I\sqcup I \rightarrow I$ } This is the spacetime in
which two disjoint intervals come together to form a single interval,
so that the spatial homology $H_0$ transitions from $\bZ^2$ to $\bZ$
as one moves past the joint at $s_0 \equiv (t_0,x_0)$. This is an example
of a ``topology changing'' spacetime, i.e., where the spatial topology
changes with time. Such spacetimes are not globally hyperbolic and
hence the analytical results of \cite{homology} are not valid in
general. Indeed, unless one admits a metric that is degenerate at $s_0$,
such a spacetime is not even causal \cite{geroch,rds,morse}. Since
acausal spacetimes cannot be discretised to obtain a causal set, any
topology changing spacetime of relevance to causal sets will have such
isolated degeneracies. On the other hand, the very nature of causal
set discretisation means that isolated points are not themselves of
relevance to the causal set, which only records the coarse grained or
relevant features of topology change. Hence there is a natural causal
set approximation to a topology changing spacetime, which is
nevertheless free of the latter's attendant pathologies.

However, the regions of the spacetime either before or after the
topology changing region are globally hyperbolic, and hence the
analytical results are valid here. We consider two different sets of
trials (a) and (b), both for $N=15,000$ element causal sets, obtained
by sprinkling into a region (a) sufficiently before the singularity,
as well as (b) sufficiently after the singularity. In both cases the
singularity is at $s_0=(0.5,0.5)$ and one thickens up to
$n_m=999$. For $t<0.5$, one is in the ``split legs'' region, with the
two strips $x\in [0,0.5]$ and $x \in [0.5, 1]$ and for $t>0.5$, one
has the single strip $x\in [0,1]$. In set (a) of the trials $t\in
[-9,1]$, so that the antichains tend to lie before the singularity,
while in set (b) $t \in [0,10]$, so that the antichains tend to lie
after the singularity. The number of trials in which $H_0=\bZ^2$
appears as the first stable region is 89 out of all 89 legitimate
trials for (a) and those for which $H_0=\bZ$ is the first stable
region is 91 out of 100 legitimate trails for (b). For 7 of the other
9 cases in (b), $H_0=\Z^2$ is the first stable region. Examining each
such case, one notices that this is because all the antichains lie in
the region before the transition. This difference between (a) and (b)
is due to the fact that the random antichain algorithm prefers
antichains that lie in the lower half of the antichain and hence the
homology of the initial region with spatial topology $I \sqcup I$
is sometimes picked up. The other two cases correspond to antichains that
are almost null and which lie above the transition region. Thus, they
have smaller cosmological scales; for one, the cosmological scale is
reached before a stable region can form, and in the other, there is an
initial stable disconnected region with $H_0=\Z^3$ and a second stable
region with $H_0=\Z$. 

\begin{figure}[ht]
\psfrag{v}{n}
\centering \resizebox{4.5in}{!}{\includegraphics{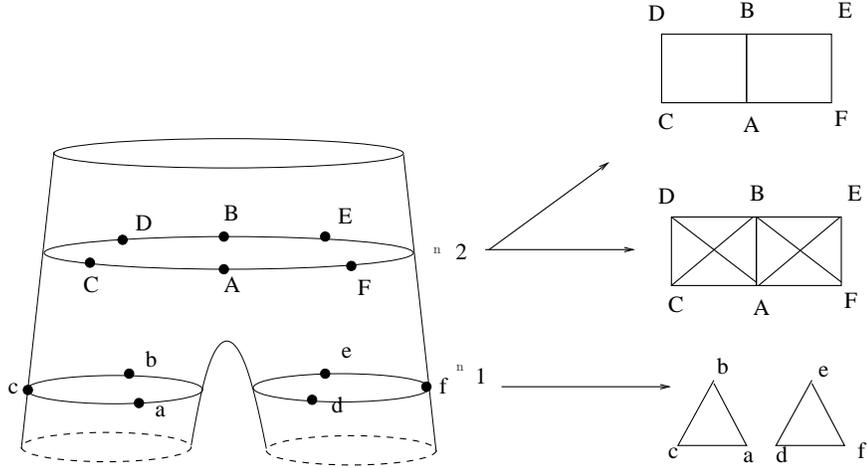}}
\caption{ The stitched trousers topology $S^1 \sqcup S^1 \rightarrow
  S^1$. Starting from a hypersurface in the region before the topology
  change with topology $S^1 \sqcup S^1$ (shown as two dashed circles)
  the correct homology $H_0=\bZ^2$, $H_1=\bZ$ for each connected
  component  is obtained for thickenings $n_1$ that lie below the convexity volume. For $n_2$
  larger than this volume, and past the singularity, depending on how
  the discretisation is done, one can get different homologies, none
  of which bear any relevance to either of the spatial
  topologies. Thus, the first nerve on the right has $H_0=\bZ$ and
  $H_1=\bZ^2$, while the second (which represents the boundaries of
  two tetrahedra joined at an edge) has $H_0=\bZ$, $H_1=0$ and
  $H_2=\bZ^2$.} \label{pants.fig}
\end{figure}

On the other hand, it is also interesting to focus our attention on
the topology changing region and to see if our trials throw any light
on it. In the simple example of the split trousers, because the only
non-trivial homology group is $H_0$, it {\it appears} that the
straddling region can actually capture the topology change. Namely,
starting with an antichain that lies in $I\sqcup I$, a thickening past
the singularity will connect the two $I$'s, so that an initial region
of $H_0=\bZ^2$ is then followed by $H_0=\bZ$. Since $H_0$ simply
measures connectedness, this is clearly not sufficient; indeed, a time
reversed case would not give rise to such a transition, since the
connectivity on the original antichain can only increase, so that
$H_0=\bZ$ even past the singularity. More generally, the details of
the discretisation can greatly influence the nerve for thickenings
that straddle the singularity, as in the example of
the stitched trousers $S^1 \sqcup S^1 \rightarrow S^1$.  Pick an
initial spatial hypersurface $S^1 \sqcup S^1$ as in Fig 
\ref{pants.fig}.  At two thickenings $n_1$ and $n_2$ one can pick
different sets of points whose shadows give us different nerves. While
the shadows from an $n_1$ less than the convexity volume will give the
correct spatial homology $H_0=\bZ^2$,  and $ H_1=\bZ$ for each connected 
component, those from $n_2$ need not bear any resemblance to either of
the two spatial homologies. Which of these if any would be picked out
consistently in a discretisation as a second stable region is unclear.

\begin{figure}[hbtp]
  \vspace{9pt}
\centerline{\hbox{ \hspace{-0.5in} 
    \epsfxsize=3.5in
    \epsffile{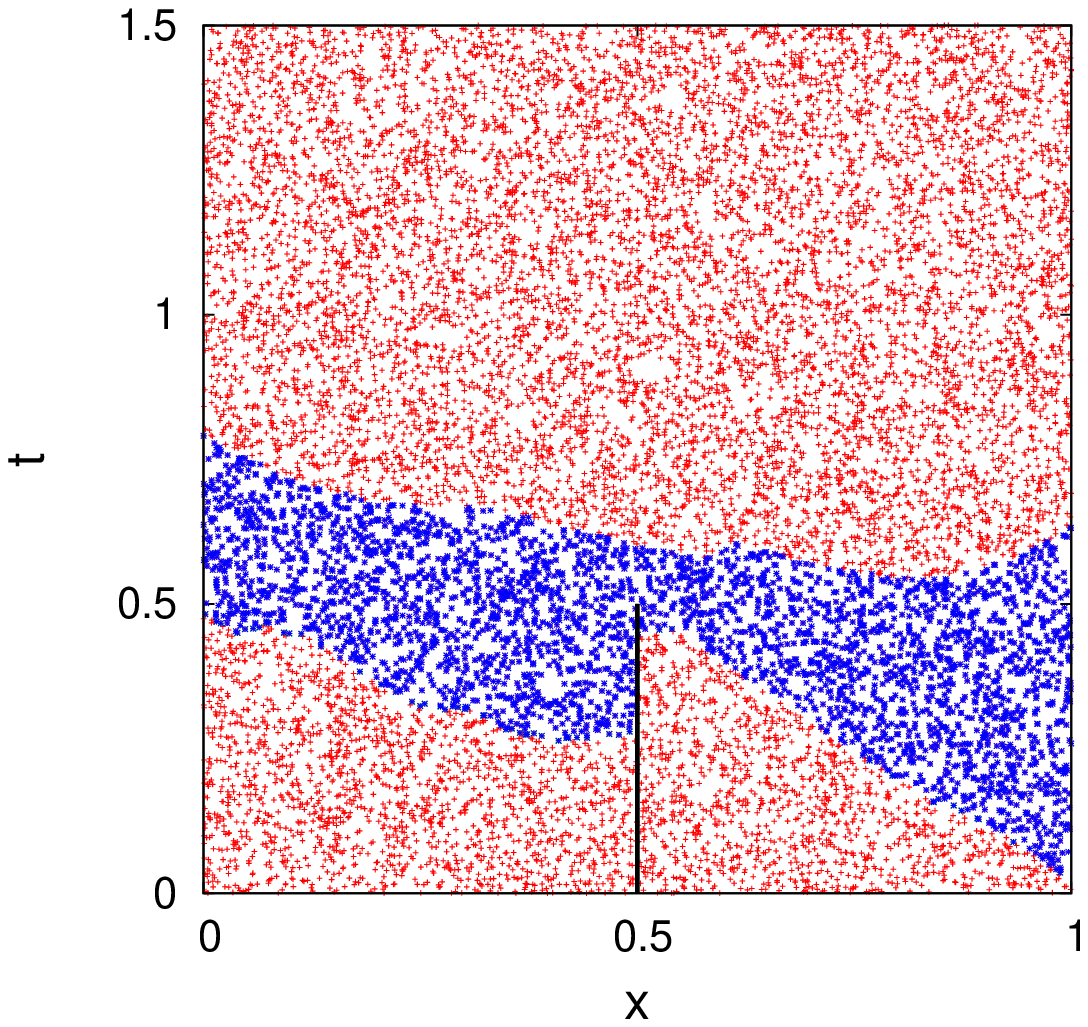} 
    \hspace{0.1in}
    \epsfxsize=3.5in
    \epsffile{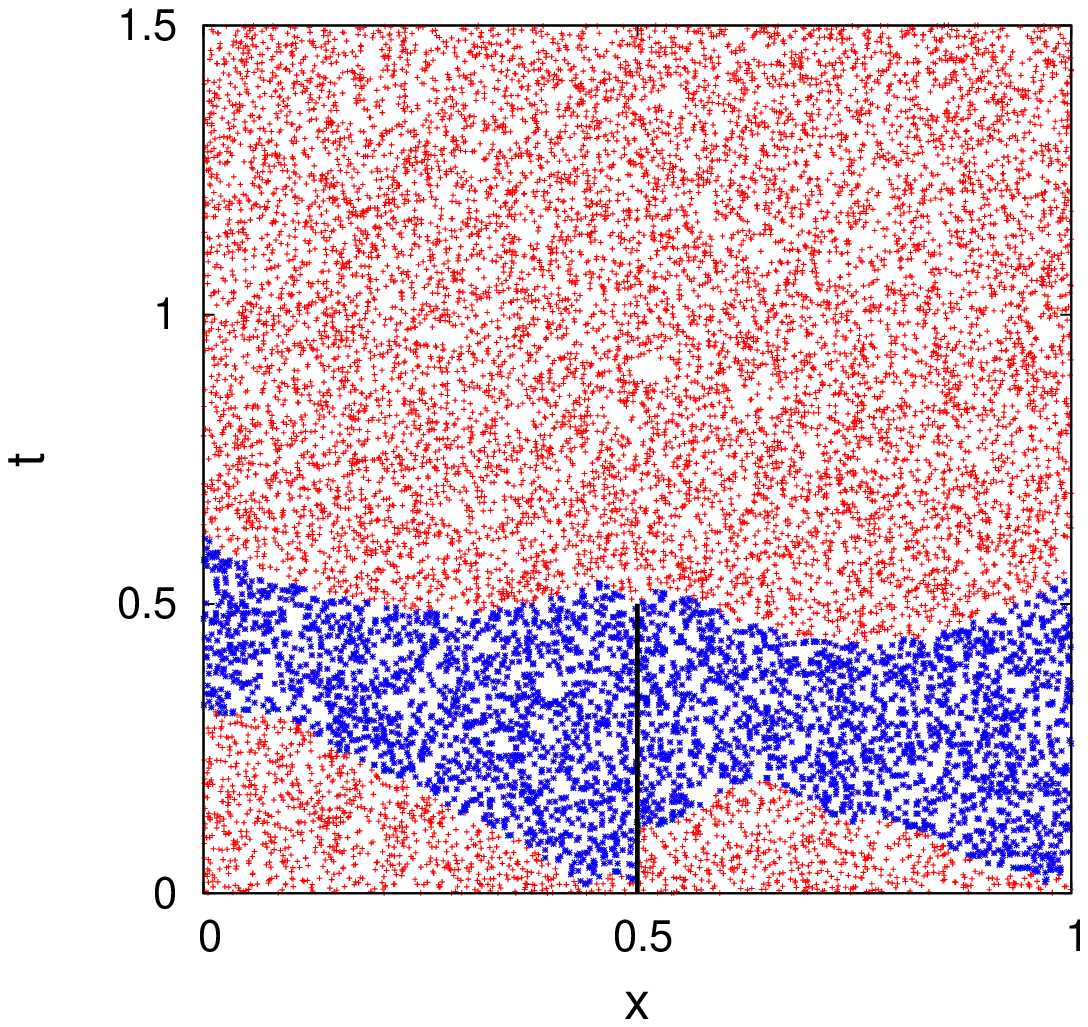}
    }
  }
  \vspace{9pt}
  \hbox
{\hspace{0.9in} 
{(a)} \hspace{3.35in} {(b)}}  
\vspace{9pt}
\centerline{\hbox{ \hspace{-0.5in} 
    \epsfxsize=3.5in
    \epsffile{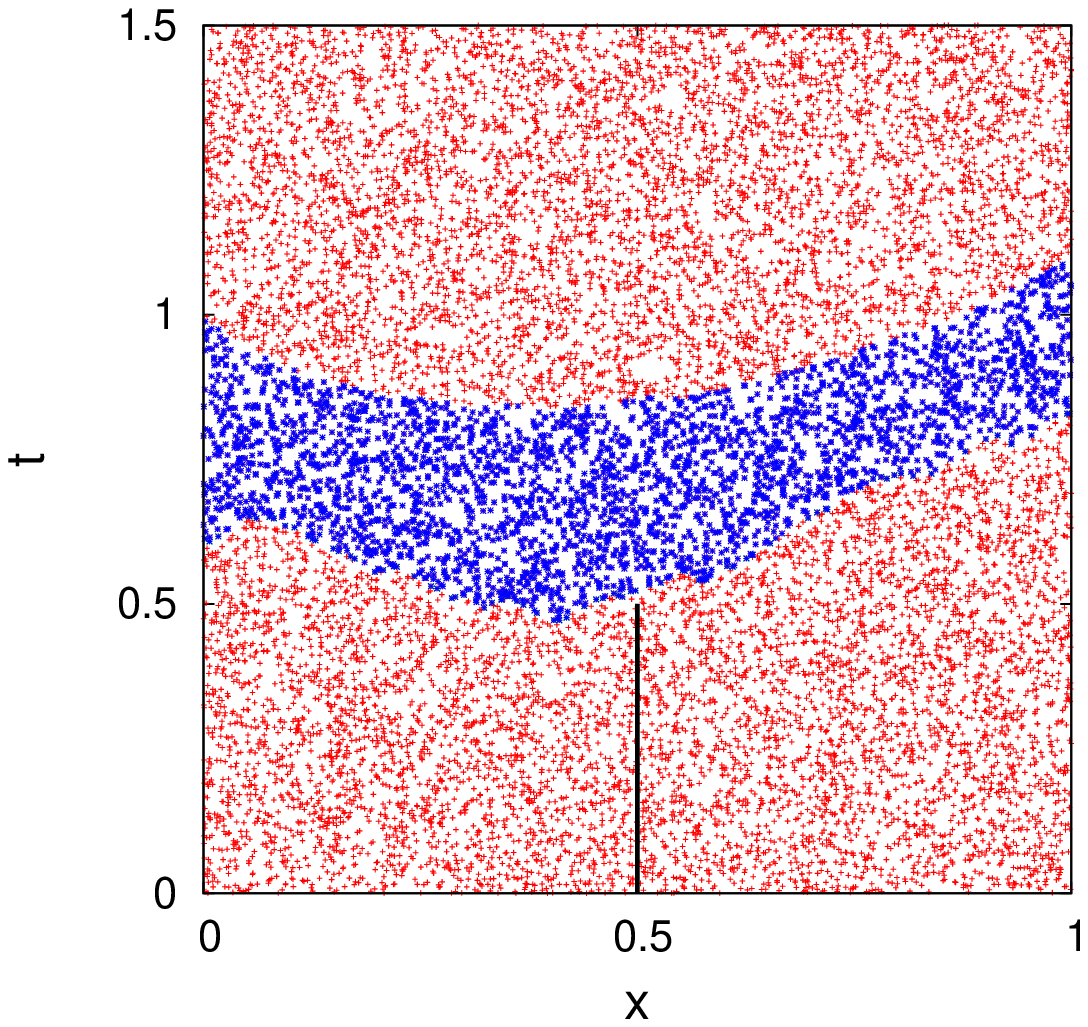} 
        }
  }
  \vspace{9pt}
  \hbox
{\hspace{2.8in} 
{(c)} } 
\caption{These are examples from the sprinklings into the trousers
  spacetime with $N=15,000 $, $x\in [0,1]$, and $t \in [0,1.5]$ (a)
  The thickened antichain straddles the region of topology change, so
  the first stable region has $H_0=\bZ^2$ followed by a stable region
  $H_0=\bZ$.  (b) This thickened antichain lies below the region of
  topology change, and so the first stable homology region with
  $H_0=\bZ^2$ is not followed by any other stable region. (c) In this
  case, the thickened antichain lies above the topology changing
  region, so the only stable homology is
  $H_0=\bZ$. } \label{trousers.fig}
\end{figure}

In order to see if we can get at least a qualitative understanding of
the region of topology change, we perform trials with an $N=15,000 $
discretisation of the split trousers with $x \in [0,1]$ and $t \in
[0,1.5]$. Again, we thicken to $n_m=999$ which is well below the
cosmological scale.  Since the topology change can no longer be
ignored, the location of the initial antichain is crucial in
deciphering the results. A careful examination of our trials shows
that for thickened antichains which straddle the region of topology
change, a first stable region with $H_0=\bZ^2$ is followed by a second
with $H_0=\bZ$.  If the antichain is chosen too close to the
transition, or well above it, only the final topology shows up as a
stable region. We get $97$ legitimate trials.  Of these, we find that
a first stable region of $H_0=\bZ^2$ followed by a second stable
region of $H_0=\bZ$, occurs in only $29$ times out of the $97$ trials.
By itself, however, $H_0=\bZ^2$ occurs as the first stable region in
$56$(including the $29$ above), while $H_0=\bZ$ occurs as the first
stable region in $32$ of them. The bias towards the former is related
to our choice of picking antichains in the lower half of the bounding
box, to avoid running into maximal elements. If instead we ask if
$H_0=\bZ^2$ or $H_0=\bZ$ occur as stable regions, with the condition
that if only the first or second appear, then they must be the first
stable region, and if they occur together they must appear one after
the other, this occurs in $87$ of the trials. In general, though, it
would seem difficult to assess manifoldlikeness for a topology
changing region, without additional restrictions on the coarse-grained
locations of the random antichains used.  Fig \ref{trousers.fig}
shows three trials in which the thickened antichain straddles three
different topological regions of the spacetime.


We should reiterate here that although the region of topology change
is not globally hyperbolic, when we {\it do} restrict to the globally
hyperbolic regions in the spacetime, as in case (a) and (b) above, we
obtain the correct continuum homology of these regions. However, given
a causal set $C$, our random antichain algorithm is not currently
suited to pick out the different globally hyperbolic regions. One way
to achieve this would be to construct a partially ordered set from the
set of all possible inextendible antichains of $C$, such that an
antichain $A$ precedes another $B$ iff no element of $B$ precedes an
element of $A$. A chain in this poset would correspond to a
``foliation'' of the causal set by inextendible antichains. Such a
foliation would not be difficult to generate, by a variety of methods.
Thickening the antichains in a foliation, one could find the stable
homology as a function of the foliation parameter, and hence isolate
the various globally hyperbolic regions in $C$ if they exist. This
procedure is computationally intensive, but can be carried out in
principle.

\subsection{3d Spacetimes} 

\begin{figure}[hbtp] 
\centerline{\hbox{ \hspace{-0.5in} 
    \epsfxsize=4.5in
    \epsffile{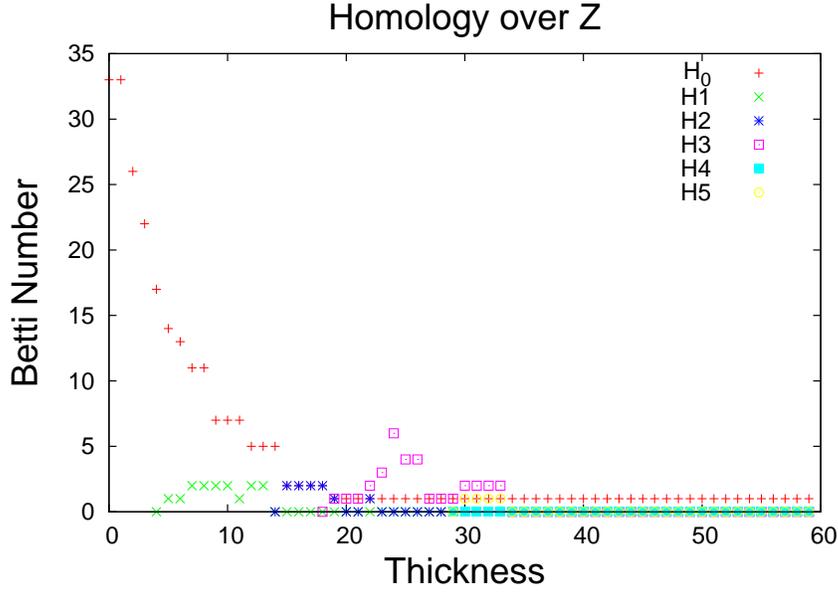} }}
\vspace{9pt}
  \hbox
{\hspace{2.9in} 
{(a)}}
\centerline{\hbox{ \hspace{-0.5in} 
    \epsfxsize=4.5in
    \epsffile{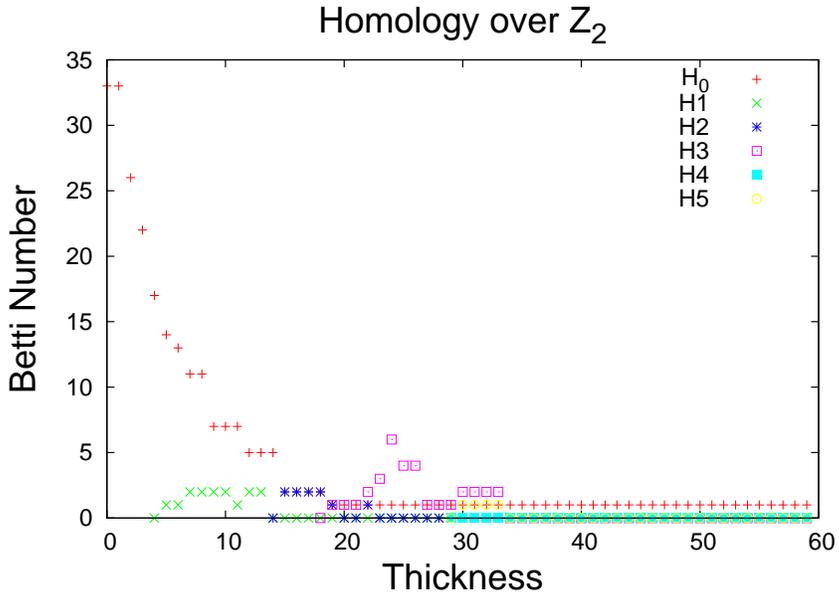}
    }
  }
  \vspace{9pt}
  \hbox
{\hspace{2.9in} 
{(b)}}  
\caption{ Comparison of homology calculations for a causal set
  approximated by a $T^2 \times I$
  spacetime, with $N=1000$. (a) uses homology on $\bZ$ and (b) uses
  the homology on $\bZ_2$. The former exhibits no torsion, so their Betti
  numbers suffice and exhibit no difference. }
\label{cheat}
\end{figure}

Although it is relatively easy to generate the nerve in higher
dimensions, the homology algorithm CHomP slows down considerably
because of the large number of high dimensional simplices that are
generated. Thus, a statistical analysis along the lines carried out
for the 2d examples is not possible, and we will use the homology
calculations to reinforce qualitatively what we have already observed
in the 2d examples. On the other hand, the 3d computations also allow
us to consider a compactified direction whose size can be varied with
respect to the discreteness scale, thus studying the effects of coarse
graining on the region of stable homology.  It helps speed up our
computations to use the preferred minimal antichain, which is what we
will do in all of the trials. For some of the examples we also
calculate homologies over $\Z_2$ instead of $\Z$, which again cuts
down the run times considerably.  This would suggest a reduction of
information, but several tests comparing the two do not find any
differences. Figures \ref{cheat} show this for a specific example.

Before we proceed with the examples it is useful to explain the choice
of simulations that we exhibit here. Although several tens of 3d
trials were started, many had to be terminated, because they could not
be run to completion within reasonable time. Estimates of run times
using different parameters were made and a process was deemed too slow
(using the RRI-AMD cluster of machines), if it took more than a month
to compute up to n=25. For low sprinkling densities, of course, the run
time is also shorter. However, the only stable homology is the trivial
one for these low density trials, so the continuum does not manifest
itself at all for the non-trivial topologies. A handful of trials
therefore remained which provided results of value. The ones we have
picked from these show something more definitive than the others. It
is important to stress that none of the others contradict the basic
hypothesis, but we do not discuss them here because they either are
not as complete or the sprinkling density is too small.

\begin{figure}[hbtp]
  \vspace{9pt}
\centerline{\hbox{ \hspace{-0.5in} 
    \epsfxsize=4.5in
    \epsffile{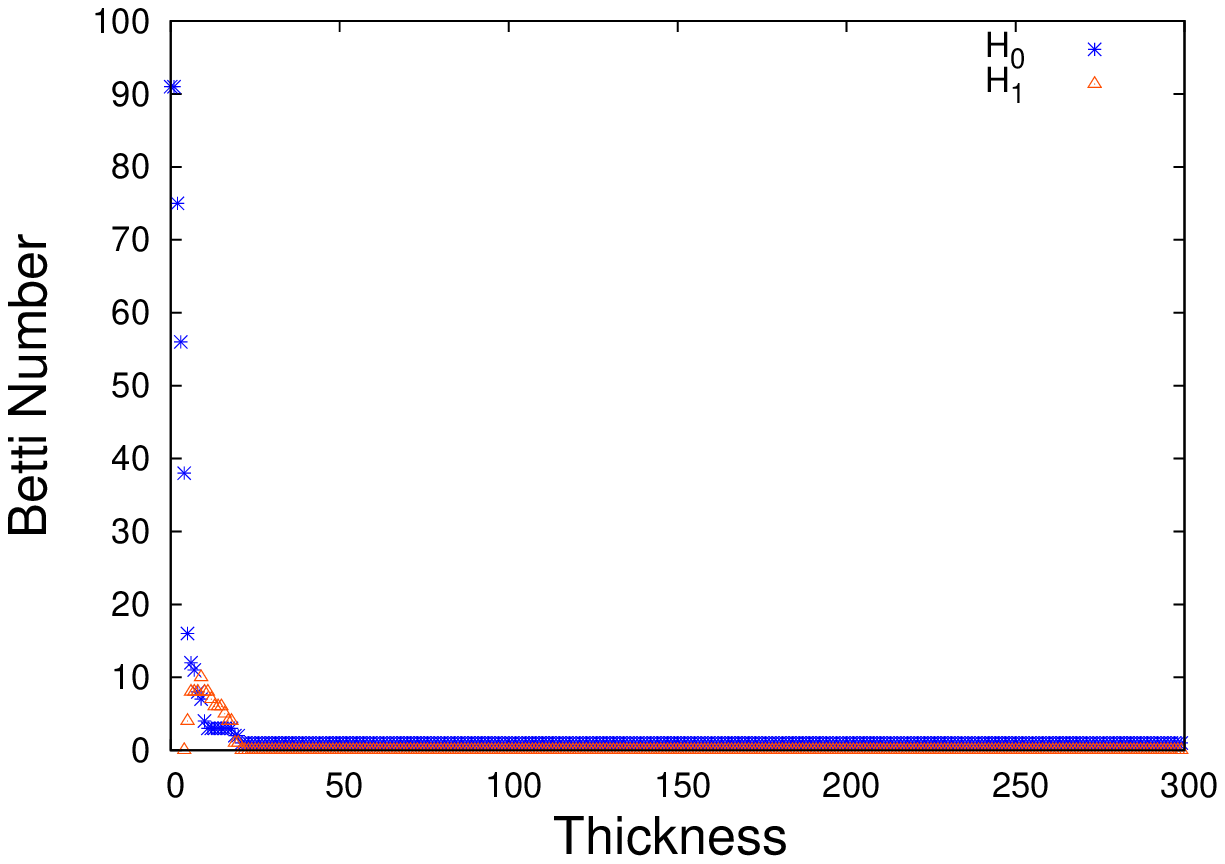}
    }}\vspace{9pt}
  \hbox
{\hspace{2.9in} 
{(a)}}
\centerline{\hbox{ \hspace{-0.5in} 
    \epsfxsize=4.5in
    \epsffile{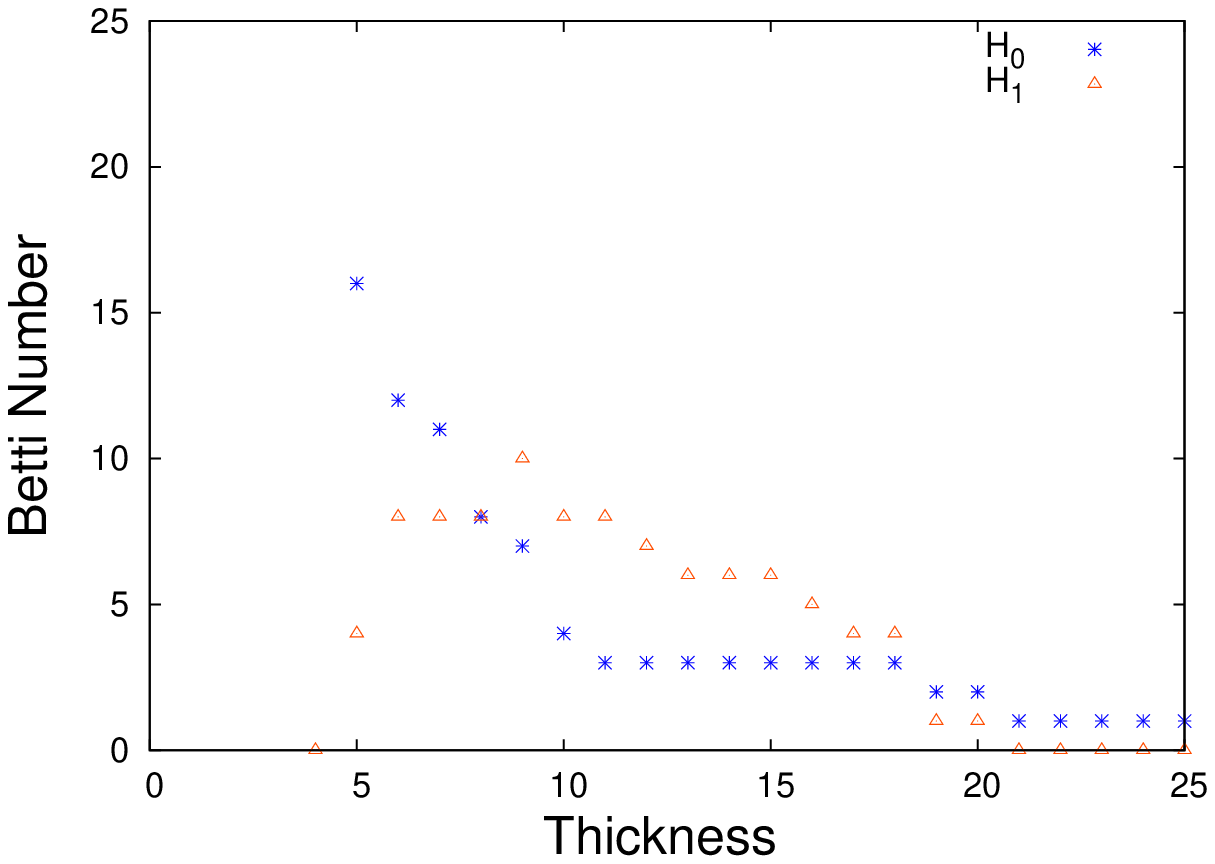}
    }
  }
  \vspace{9pt}
  \hbox
{\hspace{2.9in} 
{(b)}}  
\caption{(a) The stable homology for an $N=1024$ element causal set
  obtained from a sprinkling into a 3d interval spacetime. The first
  stable region is that of the continuum. (b) For small $n$ the
  homology is rapidly varying.} \label{IxIxI1024}
\end{figure}

\begin{figure}[hbtp]
  \vspace{9pt}
\centerline{\hbox{ \hspace{-0.5in} 
    \epsfxsize=4.5in
    \epsffile{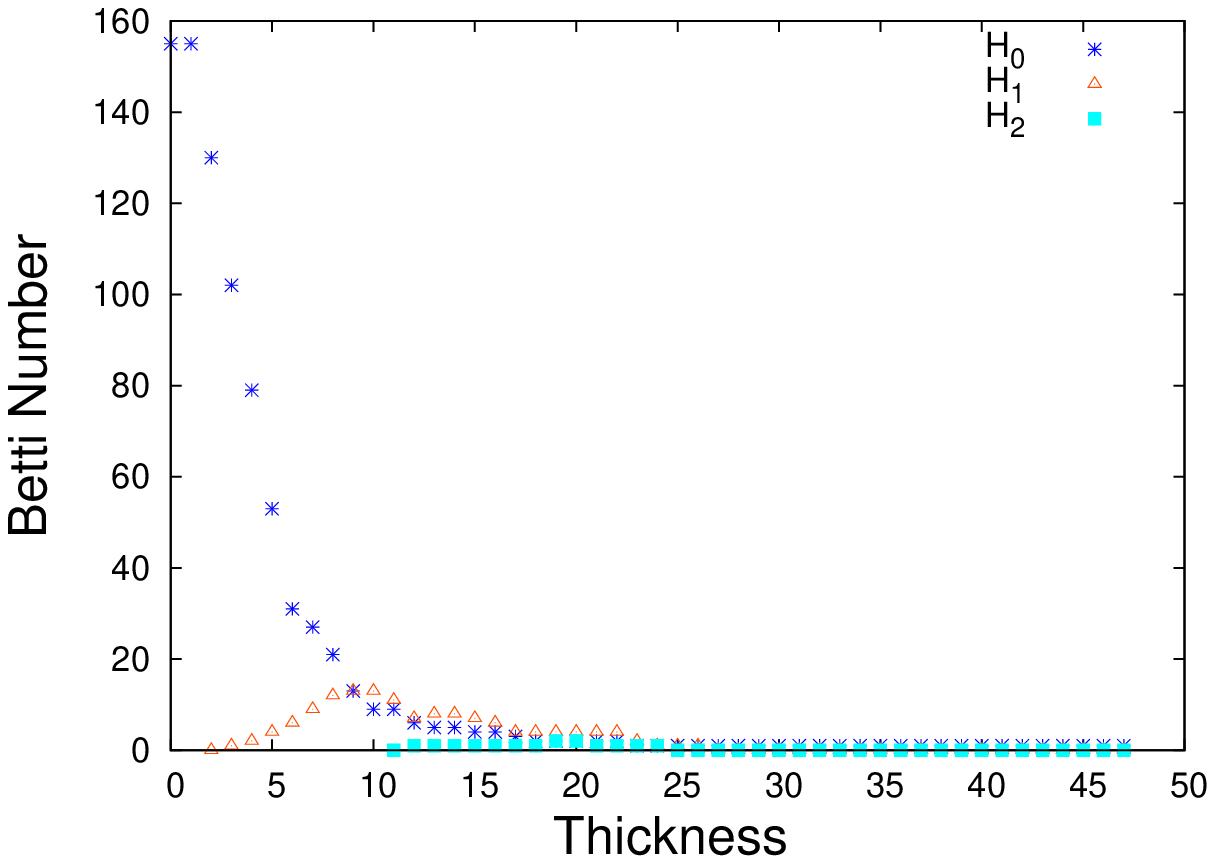}
}}\vspace{9pt}
  \hbox
{\hspace{2.9in} 
{(a)}}
\centerline{\hbox{ \hspace{-0.5in} 
    \epsfxsize=4.5in
    \epsffile{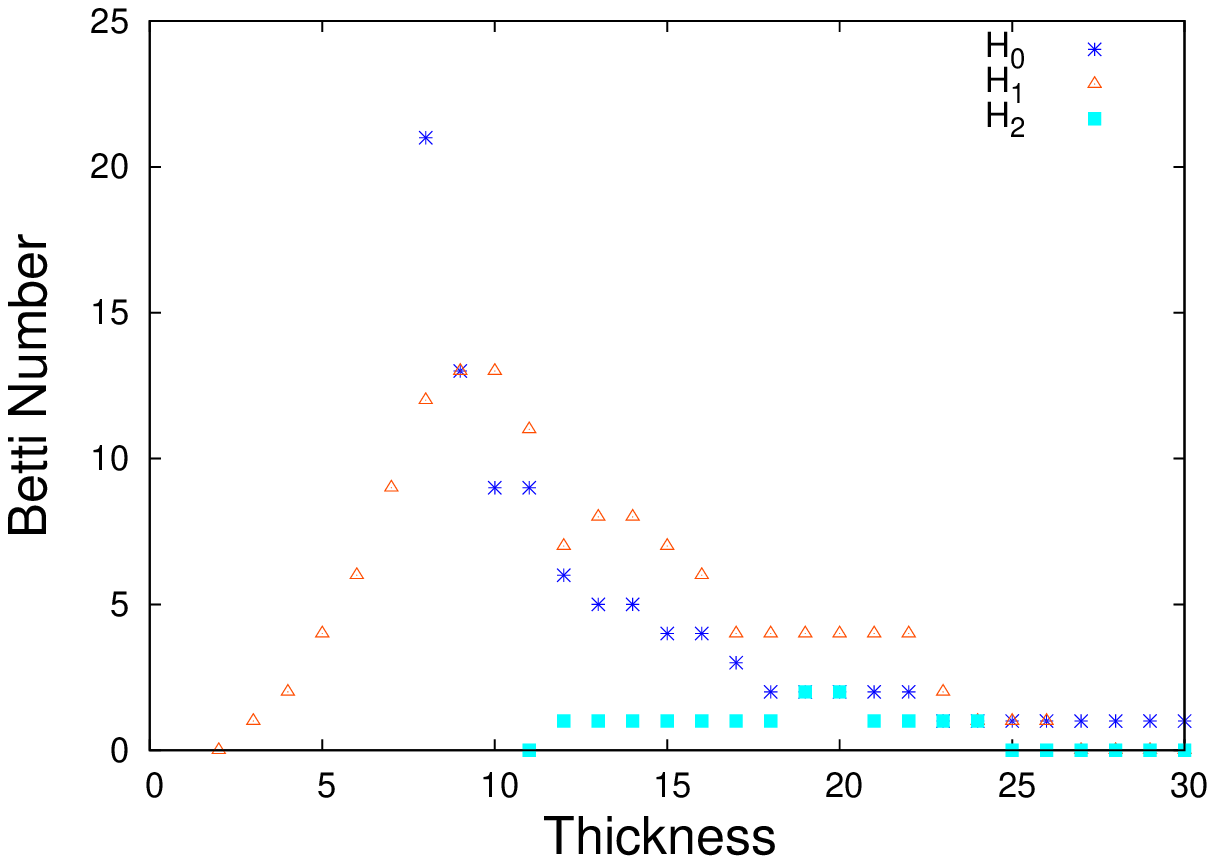}
    }
  }
  \vspace{9pt}
  \hbox
{\hspace{2.9in} 
{(b)}}  
\caption{(a) The stable homology for an $N=2048$ element causal set
  obtained from a sprinkling into a 3d interval spacetime. The
  continuum homology begins to appear as a constant homology region.
  (b) For small $n$ the homology is rapidly varying. $H_2$ is also
  non-zero in this region, unlike the $N=1024$ case.}
\label{IxIxI2048}
\end{figure}

We examined the following three spacetime topologies: 

\noindent (a) The 3d Minkowski interval spacetime with topology  $I
\times I \times I$: 
$ds^2=-dt^2+dx^2+dy^2$, $t\in [0,1]$, $x\in[0,1]$, $y\in [0,1]$.  We
calculate the homology over $\Z$ as in the 2d case, because of the
relative simplicity of the homology.  We show  in Figures \ref{IxIxI1024}
and \ref{IxIxI2048} examples with low
sprinkling densities, $N=1024$ and $N=2048$.  In both cases we see
that the first stable region is indeed the continuum one, with the
choice of mesoscale $m_s = 100$ as in the 2d case. 

\begin{figure}[hbtp]
  \vspace{9pt}
\centerline{\hbox{ \hspace{-0.5in} 
    \epsfxsize=4.5in
    \epsffile{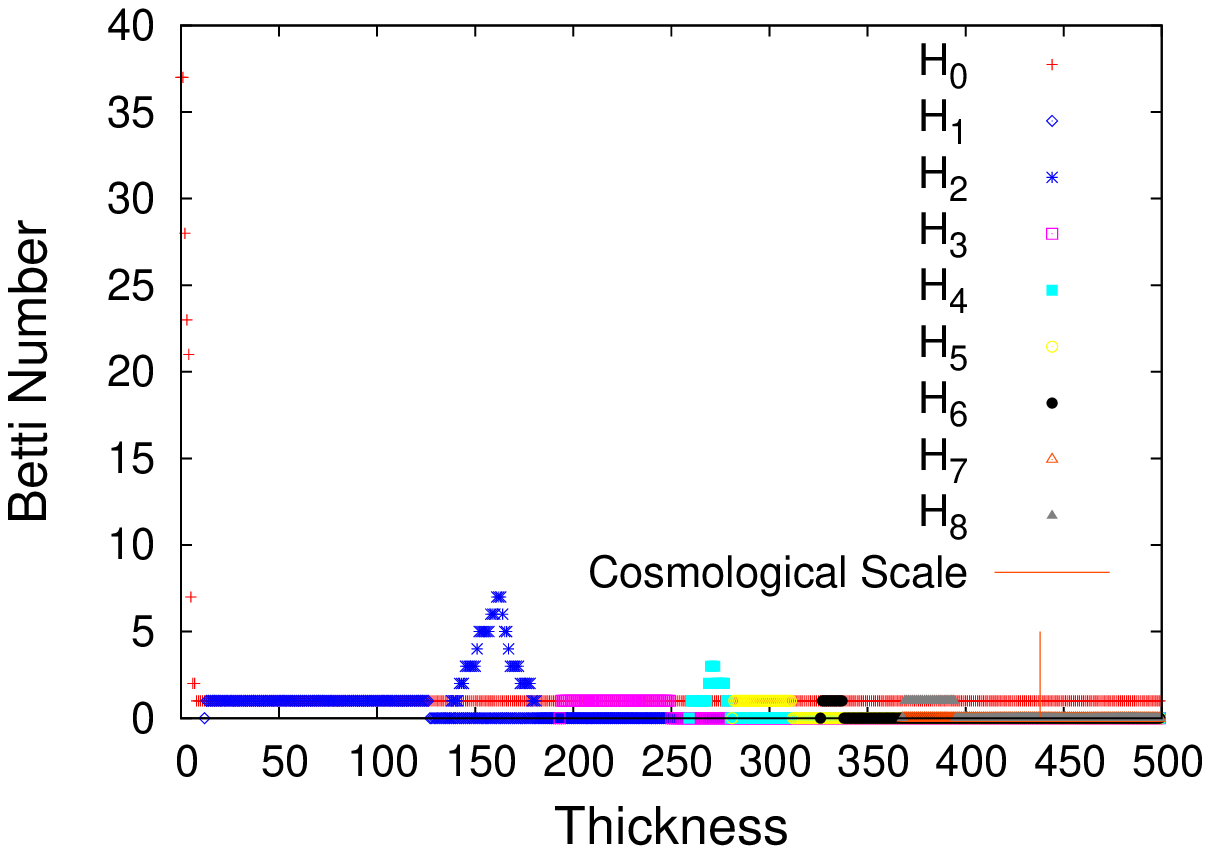}
}} \vspace{9pt}
  \hbox {\hspace{2.9in} 
{(a)}}
\centerline{\hbox{ \hspace{-0.5in}
    \epsfxsize=4.5in
    \epsffile{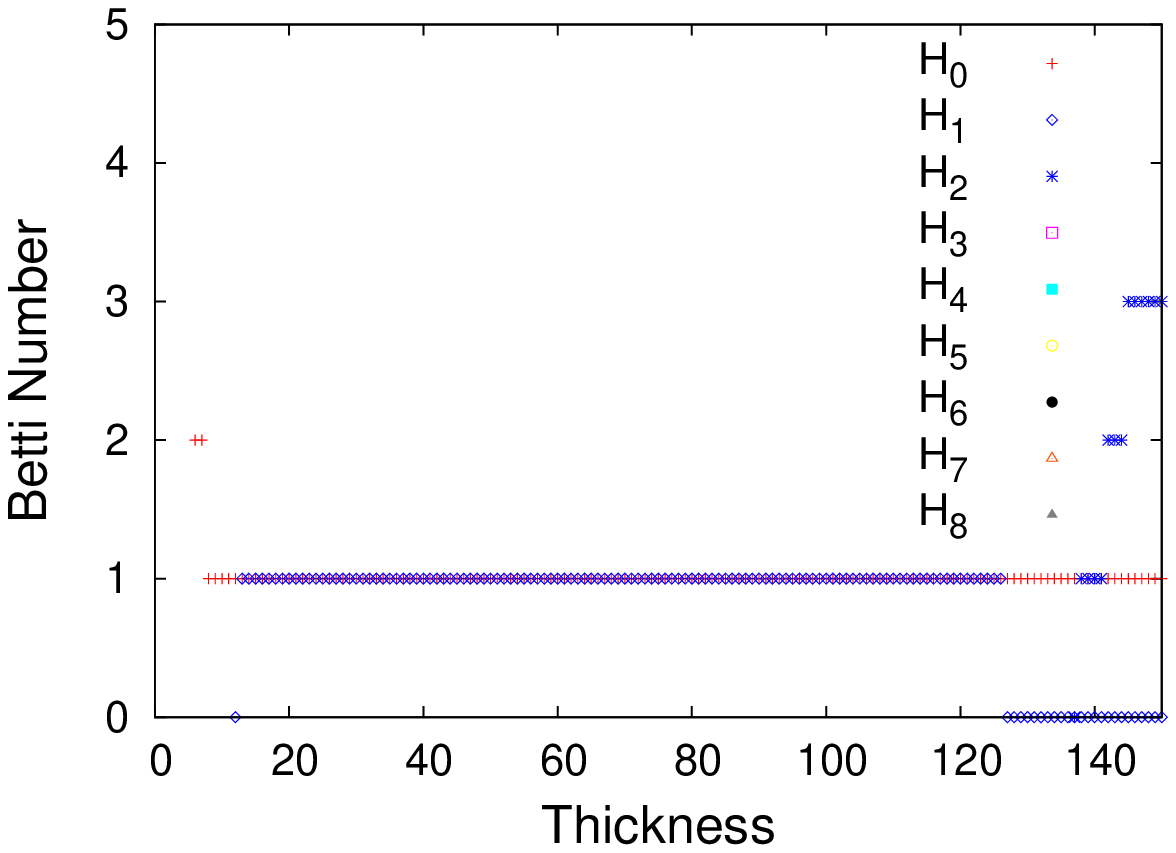}
    }
  }
  \vspace{9pt}
  \hbox
{\hspace{2.9in} 
{(b)}}  
\caption{(a) The homology for a minimal antichain for an $N=2048$
  element causal set obtained from the $S^1 \times I \times I$
  spacetime. (b) Close up of the first stable
  homology region which agrees with that of the
  continuum.}\label{intcirc} 
\end{figure}

\noindent (b) The 2d Minkowski interval spacetime with an $S^1$
compactified direction, with $t \in [0,1]$, $x_1 \in [0,1]$ and $x_2
\in [0,1]$, $x_2=0 \sim x_2=1$. The trials were carried out for
$N=1024$, $N=2048$ and $N=4096$, and the computation proceeded well
beyond the stable region. The homology over $\bZ$ was computed. In all
three cases, the first stable region is that of the continuum, i.e.,
$H_0=\bZ$, $H_1=\bZ$, again with $m_s=100$. We show one of these
examples in Fig \ref{intcirc} where $N=2048$.

\begin{figure}[hbtp]
  \vspace{9pt}
\centerline{\hbox{ \hspace{-0.5in} 
    \epsfxsize=4.5in
    \epsffile{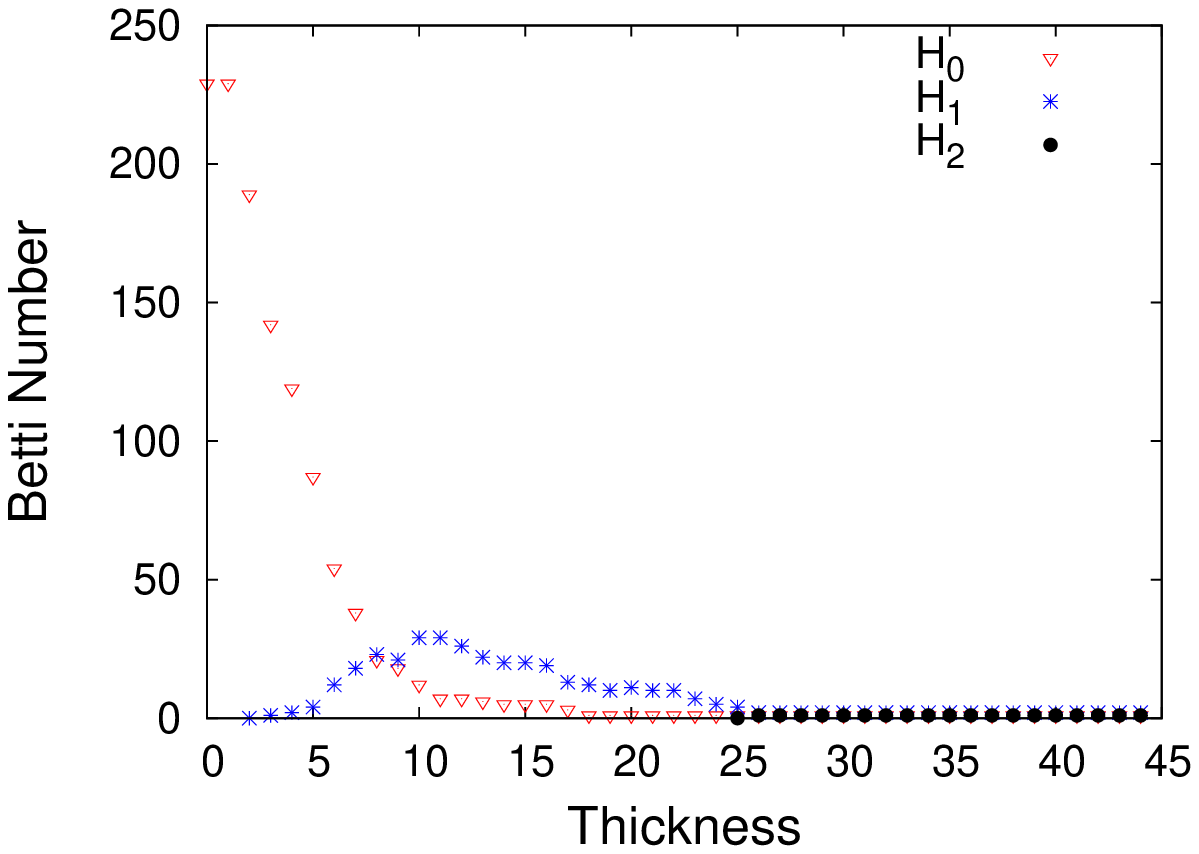}
}}  \vspace{9pt}
  \hbox
{\hspace{2.9in} 
{(a)}}    
\centerline{\hbox{ \hspace{-0.5in}
    \epsfxsize=4.5in
    \epsffile{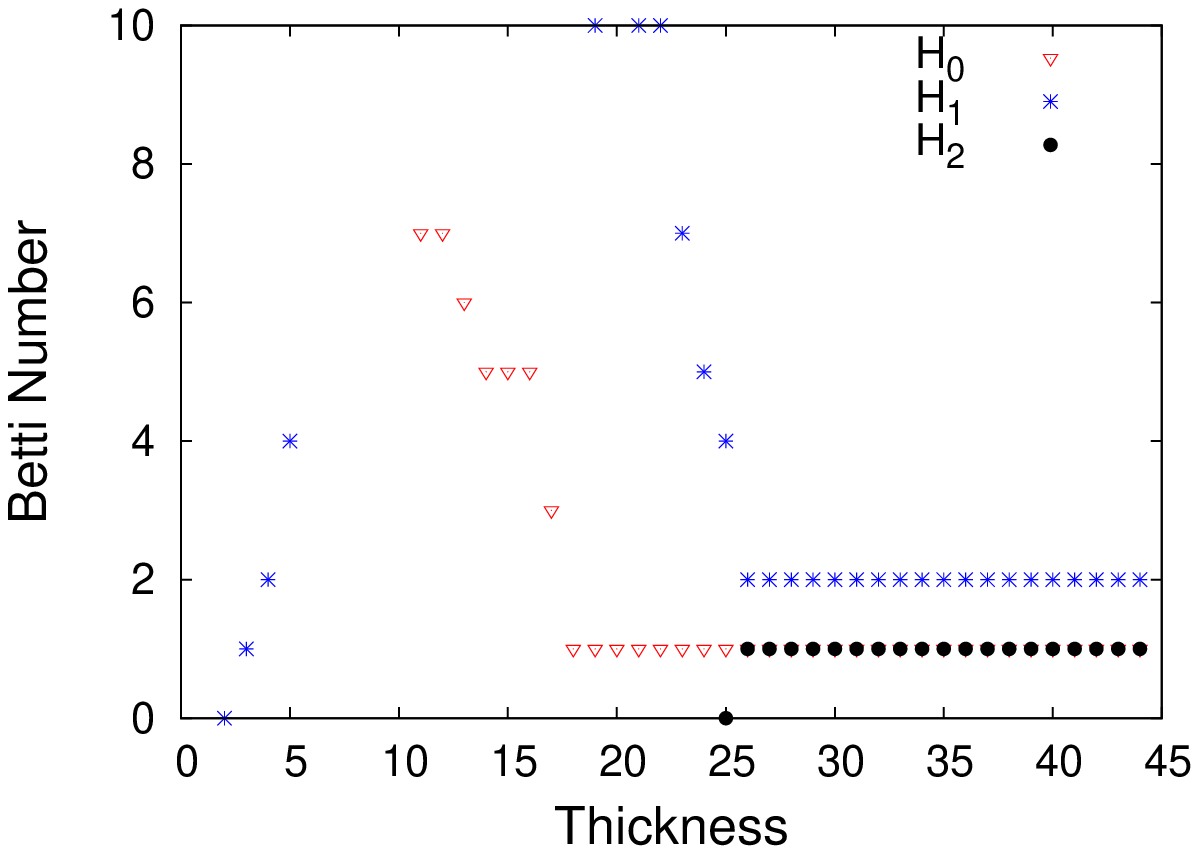}
    }
  }
  \vspace{9pt}
  \hbox
{\hspace{2.9in} 
{(b)}}
\caption{(a) An $N=4096$ causal set obtained from sprinkling into a $T^2
  \times I$ spacetime, with $a/b=1$. From $n=26$ to $n=44$ the
  homology is that of $T^2$ with $H_0=\bZ$, $H_1=\bZ^2$ and
  $H_2=\bZ$. The computation was stopped before the end of this region
  could be found. (b) A closer look.}  
\label{T2xI1}
\end{figure}

\begin{figure}[hbtp]
  \vspace{9pt}
\centerline{\hbox{ \hspace{-0.5in} 
    \epsfxsize=4.5in
    \epsffile{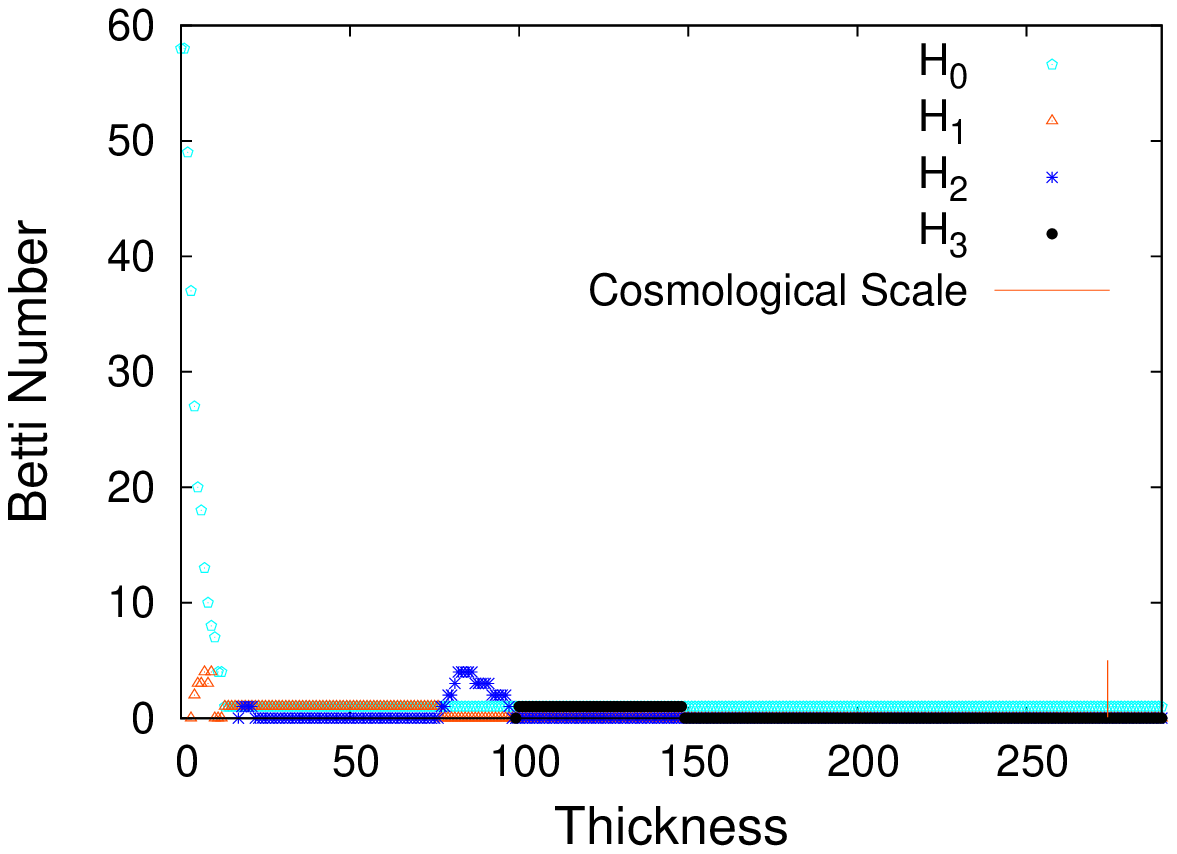}
}} \vspace{9pt}\hbox 
{\hspace{2.9in} 
{(a)}}   
\centerline{\hbox{ \hspace{-0.5in} 
    \epsfxsize=4.5in
    \epsffile{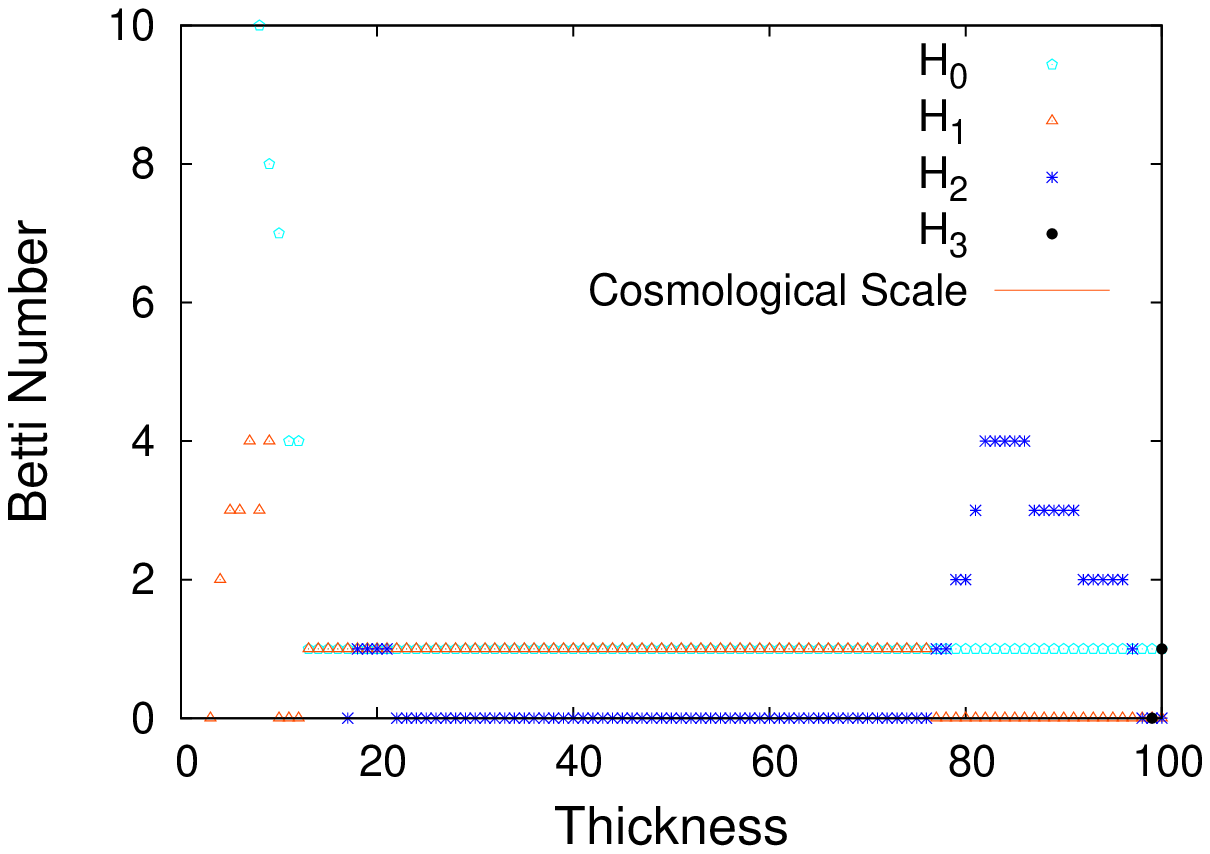}
    }
  }
  \vspace{9pt}\hbox 
{\hspace{2.9in} 
{(b)}}  
\caption{An $N=1024$ element causal set sprinkled into the $T^2 \times
  I $ spacetime  with aspect ratio $a/b=0.25$. The 2d
  homology groups $H_0=\bZ$, $H_1=\bZ$ appears as the first stable
  region. The 3d homology is not captured even fleetingly.} 
\label{T2xI.25}
\end{figure}

\noindent (c) The spacetimes with topology $T^2 \times I$:
$ds^2=-dt^2+dx^2+dy^2$, $t\in [0,1]$, $x\in[0,a]$ with $x=0 \sim x=a$,
$y\in [0,b]$ with $y=0 \sim y=b$. Trials were carried out for (i)
$a=b=1$ (Fig \ref{T2xI1}) and (ii) $a=0.25$, $b=1$ (Fig
\ref{T2xI.25}). In all cases, the homology over $\bZ_2$ was
computed. In (i) the simulations gave the best results for an $N=4096$
causal set. In this case, the 3d homology begins to appear as a
constant homology region, but the simulations could not be run beyond
this stage. For lower densities $N=1024$ and $N=2048$, these continuum
homology regions appear fleetingly, and hence are not stable.  For
(ii), trials were carried out for $N=1024$ and $N=2048$ and for both
the 2d cylinder spacetime homology $H_0=\bZ_2$ and $H_1=\bZ_2$ appears
as the first stable region, again with $m_s=100$.  The 3d region does
not appear even fleetingly in these cases. The low sprinkling
densities mean that the small compactified direction $S^1$ does not
appear as a continuum feature in the causal set.


\section{Testing for Manifoldlikeness} 
\label{testman} 

We now consider two classes of causal sets neither of which is obtained
via a causal set discretisation of a spacetime. The first class
contains causal sets that are obtained from a particular classical
growth dynamics called ``transitive percolation''. The second is a 
class of regular or crystalline causal sets, the ``tower of
crowns''. Since they do not arise from a spacetime discretisation, they are
ideal candidates to put through our test for manifoldlikeness.

\subsection{Transitive Percolation}
As a precursor to understanding the quantum dynamics of causal sets, a
class of classical sequential growth dynamics for causal sets was
constructed in \cite{csg}. One starts with a single element and adds
new elements one at a time such that at stage $n+1$ the new element
cannot be added to the past of any element in the $n$-element causal
set obtained at the previous stage.  The process is Markovian and is
required to satisfy the physical criteria of label invariance and a
``Bell-causality'' condition \cite{csg}. These criteria restrict the dynamics to
a class of ``general percolation'' models studied in detail in
\cite{csg,rg,dm}, of which 
transitive percolation is of particular interest.  In
\cite{rg,Ashone,Ashtwo} it was shown that in a cosmological context,
the causal set can undergo a sequence of bounces, with each era between
bounces determined by a different generalised percolation dynamics. At
late times, it was shown that the parameters ``flow'' to that of
transitive percolation. Hence it is of interest to see if a typical
causal set obtained via transitive percolation has any manifoldlike
characteristics.

The transitive percolation dynamics is determined by a single
parameter $0\leq p \leq 1 $ which gives the probability for a new
element at stage $n$ to be linked to an element of the $n$-element
causal set obtained at stage $n-1$. Thus, starting from a single
element, the probability to get the 2-element chain is $p$ and the
2-element antichain is $1-p$, that of a 3-chain is $p^2$, that of a
3-antichain $(1-p)^3$, etc.

We run our trials for the following three choices of $p=0.05, 0.045,
0.04$ for $N=4000$ element causal sets. The reasons for these choices
are that (i) for larger $p$, the cosmological scale appears very
quickly, and our trials suggest that there are typically no stable
regions at all before the cosmological scale \footnote{For $p=0.1$,
  for example, out of $100$ trials, only $6$ had stable regions.}  and
(ii) for smaller $p$, the run-times become too large to perform a
suitable statistical sampling.  However, it is known that the number
of elements in a largest inextendible antichain, or ``width'', of a
causal set generated by transitive percolation $\sim 1/p$.
Thus, the transitive percolated causal sets we consider are quite
narrow, with a width of no more than $\sim 1/.04 = 25$.  Therefore in
this sense we are only able to access relatively small transitive
percolated causal sets computationally, and so may not be surprised
that they have trouble producing regions of stable homology beyond the
mesoscale, but before the cosmic scale sets in.

For each case, we perform $100$ trials.  For $p=0.05$, all the trials
are legitimate, although $49$ of these have no stable
regions. A case by case analysis reveals that this is because the
cosmological scale is very small for each of these trials.  We find
that $44$ of the trials have $H_0=\bZ$ as the first stable region,
$7$ with $H_0=\bZ^2$, the other homology groups all being
trivial. In none of these cases is there more than one region of
stable homology before the cosmological scale.

For $p=0.045$, again all $100$ trials are legitimate. We find that
$62$ of the trials have $H_0=\bZ$ as the first stable region, $16$
with $H_0=\bZ^2$, the other homology groups all being trivial. In
addition, $2$ of these have first stable region $H_0=\Z^2$ and second
stable region $H_0=\Z$, while $22$ have no stable regions.

\begin{figure}[ht]
\centering \resizebox{4in}{!}{\includegraphics{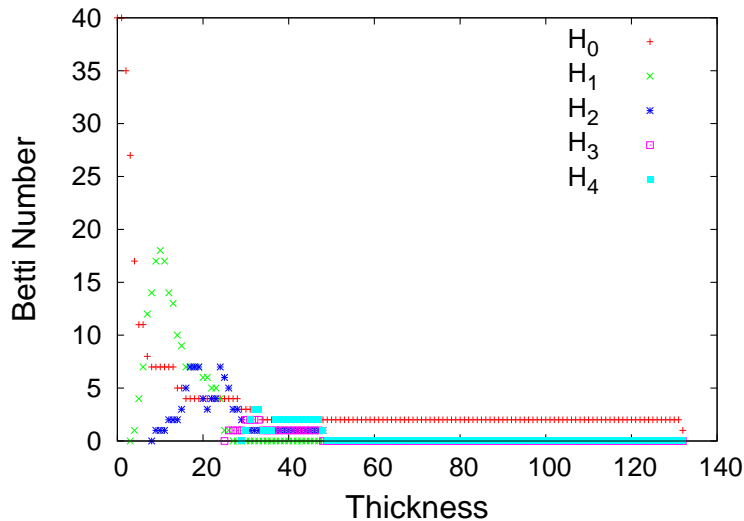}}
\caption{\small A sample of the homology groups for $p=0.04$
  transitive percolation with $N=4000$. The thickness parameter runs
  all the way to the cosmological scale. } 
\label{TP.04homs}
\end{figure}

For $p=0.04$, we find that now only $53$ of the $100$ legitimate
trials have $H_0=\bZ$ as the first stable region, $30$ have
$H_0=\Z^2$, and only $1$ of these $30$ has the first stable region
with $H_0=\bZ^2$ followed by a second stable region with
$H_0=\bZ$. There is only one other case with stable first stable
homology $H_0=\Z^3$ followed by a second stable region of
$H_0=\Z^2$. There are $16$ trials with no stable regions.  Fig
\ref{TP.04homs} shows an arbitrary sample of the homology groups which
arise from an inextendible antichain from transitive percolation at
$p=0.04$. The $H_0 = \Z^2$ homology occurs as a stable region for this
antichain.  Some of the higher homology groups, up to $H_4$, become
non-trivial before the first stable region begins, thus contributing
to the ``discreteness noise''. This is a feature we do not see in the
causal sets obtained from 2d discretisations.

Thus, it appears that for these values of $p$, the causal sets do not
pass our test for manifoldlikeness with sufficient confidence,
although the appearance of stable regions, along with an initial
period of instability seem suggestive.  It is tempting to conclude
that these results are compatible with the split trousers topology in
the region of topology change. There are definitely similarities,
especially with the $p=0.45$ case. If the existence of a second stable
region before the cosmological scale is an indicator of topology
change on the other hand, there are only $6$ instances of a $H_0=\Z^2$
to $H_0 = \Z$ transition. If the comparison to the inverted trousers
had to be made, then as argued earlier, there can be no antichains for
which such a transition can be made.  However, it does seem {\it
  plausible} that one can ``tailor'' a trousers spacetime to suit the
outcome of this trial.  On the other hand, it is important to be
cautious in this comparison. Neither the trousers nor these cases of
transitive percolation satisfy our test of manifoldlikeness. Further
work on stable homology in topology changing spacetimes is required
for us to conclude that this similarity is more than incidental.


\subsection{Tower of Crowns}  
\label{crowns} 

It is useful to consider as an  example a regular causal set, the tower
of crowns, which embeds into the 2d cylinder spacetime, though not
faithfully. Its regularity means that it has regions of stable
homology. In this case, this homology is that of $S^1$, and hence it
seems to pass one of the tests for manifoldlikeness. However, its
regularity also means that for small thickenings, the fluctuations are
not sufficiently uncorrelated. The simplicity of the causal set allows
us to examine it analytically. 

\begin{figure}[hbtp]
  \vspace{9pt}
\centerline{\hbox{ \hspace{-0.5in} 
    \epsfxsize=2.5in
    \epsffile{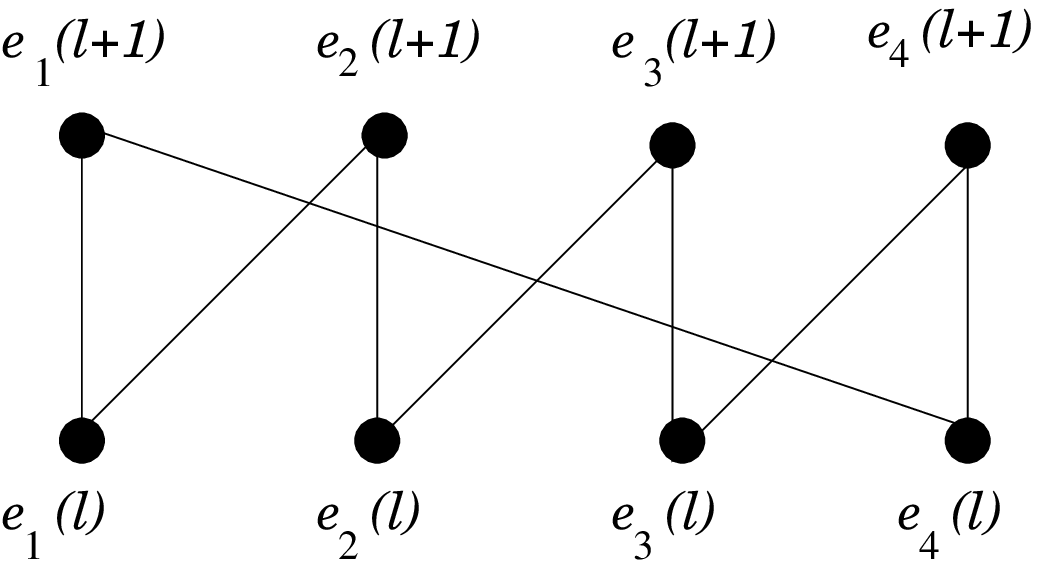} 
    \hspace{0.25in}
    \epsfxsize=2.5in
    \epsffile{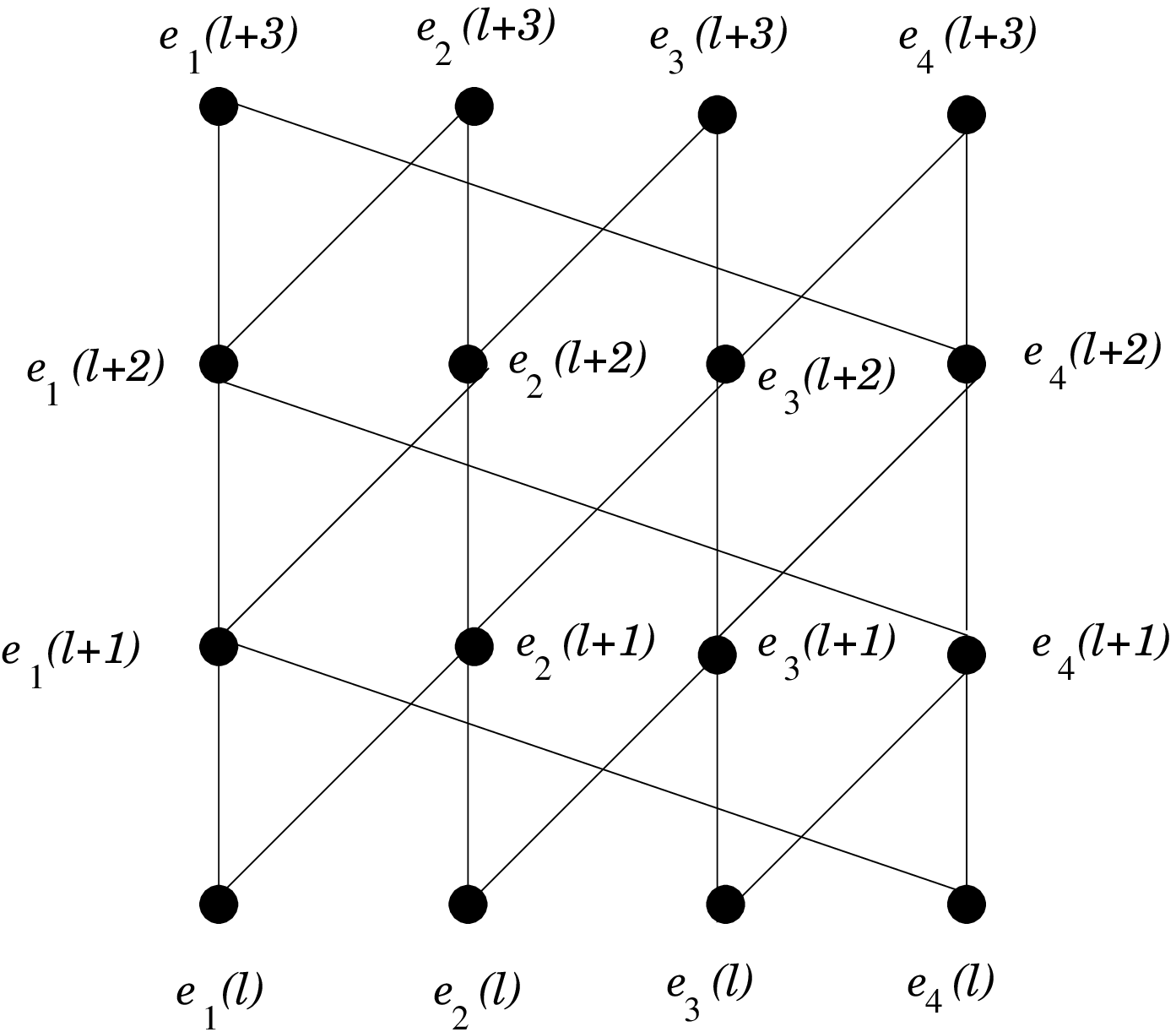}
    }
  }
  \vspace{9pt}
  \hbox
{\hspace{1.25in} 
{(a)} \hspace{2.5in} {(b)}}  
\caption{(a) An $m=4$ crown with $8$ elements. (b) A tower of $m=4$
  crowns.} \label{crown}
\end{figure}

A crown is a 2 layer causal set, with $m>2$ elements in each layer. It
has a natural labelling in terms of which the order relations can be
expressed. Namely, if $\{e_1(0), e_2(0) \ldots e_m(0)\} $ are the elements in
the bottom layer and $\{e_1(1), e_2(1) \ldots e_m(1)\} $ in the top
layer, then $e_i(0) \prec e_i(1)$ and $e_i(0) \prec e_{i+1}(1)$ with
the labels $i \equiv i
\, {\mathrm { mod}} \, m $ (Fig \ref{crowns}(a)). Stacking an infinite number of
m-crowns one on top of the other gives us the tower of crowns (Fig
\ref{crown}(b)). It has a preferred foliation for which every level
$l$ (measured from some fiducial crown) contains $m$ elements, $\{
e_i(l)\} $, where $i=1, \ldots m$.\footnote{The numeric labels used in the
  random antichain algorithm are ordered by layer, i.e.\ elements $0 \ldots m-1$
  are in the bottom layer, $m \ldots 2m-1$ in the second, etc.} The level $l+1$ elements are
related to the level $l$ elements as follows: $e_i(l+1) \succ e_i(l)$
and $e_{i+1}(l+1) \succ e_i(l)$. The transitive closure of this gives
us the causal set. We will only use this preferred foliation for our
choice of inextendible antichains, to make the discussion simple. 

\begin{figure}[hbtp]
  \vspace{9pt}
\centerline{\hbox{ \hspace{-0.5in} 
    \epsfxsize=2in
    \epsffile{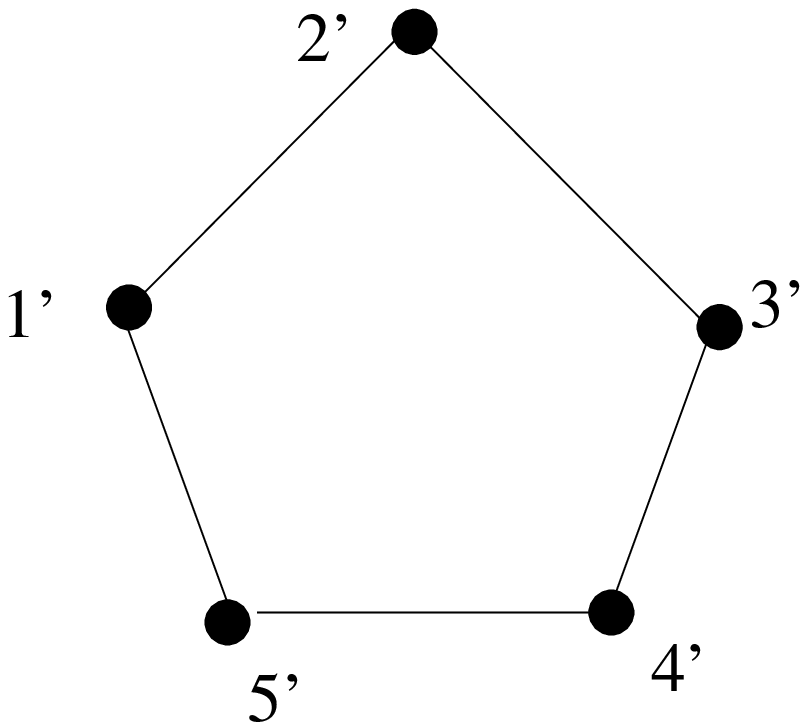} 
    \hspace{0.25in}
    \epsfxsize=3in
    \epsffile{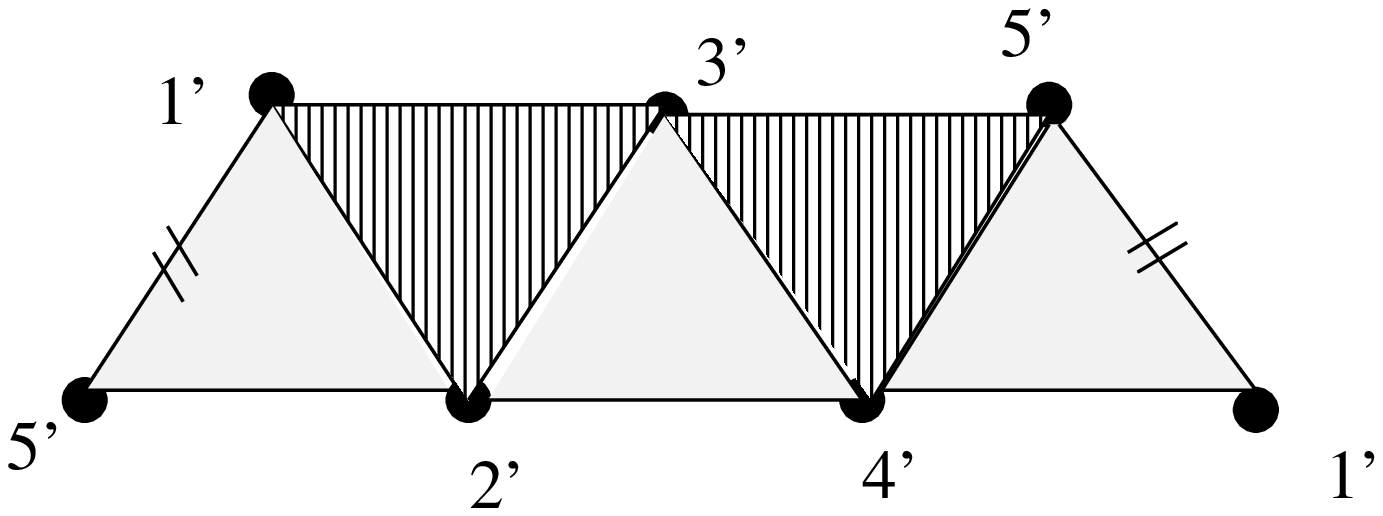}
    }
  }
  \vspace{9pt}
  \hbox
{\hspace{1in} 
{(a)} \hspace{2.4in} {(b)}}  
\caption{The nerves of (a) $T_1(A)$ and (b) $T_2(A)$ for $m=5$. } \label{cycle} 
\end{figure}

Let us start with the $l=0$ level antichain $A$ and thicken to
$T_l(A)$.  Because of the regularity of the crown poset, this level
thickening is closely related to volume thickening. Thus, $l=1$
corresponds to a volume thickening $n=2$, level $l=2$ is $n=5$, $l=3$
to $n=9$. All intermediate values of $n$ are identical to the corresponding lower values. The set of
shadows in $T_1(A)$ are then $\{S_i(1)\equiv \{e_i(0), e_{i-1}(0)
\}\}$. $S_{ij}(1) \equiv S_i(1)\cap S_j(1) \neq \emptyset$ only for
$j=i+1$ and $j=i-1$, with the intersections being $e_i(0)$ and
$e_{i-1}(0)$ respectively. Thus, 3-way and higher intersections
vanish.  For $m \geq 3$, the nerve $\cN(1)$ associated with $T_1(A)$ is
therefore a cycle of 1-simplices, with $m$-vertices and $m$-edges as
shown in Fig \ref{cycle}(a). Thus $H_0=\bZ$ and $H_1= \bZ$,
i.e. $\cN(1)$ is homological to the circle. At level $l=2$, the shadows
are $\{S_i(2) \equiv \{e_i(0), e_{i-1}(0), e_{i-2}(0) \} \}$, and the
largest intersections are 3-way, $S_{ijk}(2) \equiv S_i(2) \cap S_j(2)
\cap S_k(2)$, with (i) $j=i+1$, $k=i+2$, $S_{ijk}(2)=e_i(0)$, (ii)
$j=i+1$, $k=i-1$ $S_{ijk}(2)=e_{i-1}(0)$ or (iii) $j=i-1$, $k=i-2$,
$S_{ijk}(2)=e_{i-2}(0)$.  This gives a nerve simplicial complex $\cN(2)$
made up of $2$-simplices (Fig \ref{cycle}(b)), which for $m \geq 5$ is
again homological to the circle. Note that for $m=4$, the non-zero
homology groups are $H_0=\bZ$, and $H_2= \bZ$, which is homological to
the sphere and not the circle.

At level $l$, therefore, the shadows are $\{S_i(l) \equiv \{e_i(0),
e_{i-1}(0), \ldots e_{i-l}(0)\} \}$, and one has non-trivial $l+1$ way
intersections $S_{i_1i_2\ldots i_{l+1}}(l)$ for a consecutive set
$\{i_1, i_2 \ldots i_{l+1} \}$ thus giving rise to a higher dimension
nerve simplicial complex. Now, as long as the shadows are small enough
as in the example in Fig \ref{cycle}(b), these higher dimensional
nerves are homological to the circle. Small enough is determined by
taking recourse to the structure of $\cN(1)$ which resembles a circle,
for $m \geq 3$, and provides a nearest neighbour for every element in
$A$. Let us make this more precise. 

Let $N^\ast(1)$ be the dual complex of $\cN(1)$ formed by taking
vertices to edges and vice versa. The vertices of $N^\ast(1)$ are
therefore the elements of $A$, and the edges, represent the level 1
cover of $A$, i.e. the elements of level 1 whose shadows cover $A$.
Then there exists homotopy preserving maps $\Xi: N^\ast(1) \rightarrow S^1$
such that $p_i=\Xi(e_i(0))$ are a distinct set of $m$ points on $S^1$
and the edges between $e_i(0)$ and $e_{i+1}(0)$ map to the connecting
curve from $p_i$ to $p_{i+1}$ in $S^1$. One thus obtains from this a
covering of $S^1$ by closed intervals. These can be extended trivially
to  open intervals, since $\Xi$ is only required to be homotopy
preserving. 

Consider the two shadows $S_i(l)$ and $S_{i-l}(l)$ whose intersection
includes $e_{i-1}(0)$. If $i-2l \, \mathrm{mod} \, m \leq i \,
\mathrm{mod} \, m$, however, then $e_{i}(0)$ also belongs to the
intersection. However, for $0 < l < m $, $e_{i}(0)$ and $e_{i-1}(0)$
are not nearest neighbours on $N^\ast(1)$. Using the De Rham-Weil
theorem \cite{derham-weil} we see that the related covering of $S^1$
has intersections which are not themselves connected, and hence its
nerve need not reproduce the homology of the circle. On the other
hand, if $2l < m $, all intersections are homotopically trivial, and
so the nerve $\cN(l)$ is homotopic to $S^1$.

Thus, for $m$ sufficiently large, one obtains a large range of
thicknesses $1 \leq l \leq [m/2]$ for which the homology is
stable. However, this stability sets in at the discreteness scale,
unlike the case of manifoldlike causal sets, where $H_0$ rapidly
decreases. If it were possible to isolate this class of antichains in
the tower of crowns poset without any a priori knowledge of its
structure, it would be clear that the causal set does not pass our 
test of manifoldlikeness.

However, our numerical test for manifoldlikeness uses random
antichains and we do not look for the existence of special antichains
from which to start the thickening. For a randomly chosen antichain it
will generically be true that the homology changes for small
$n$. Because of the crystalline structure, the number of inextendible
antichains which have a distinct behaviour (i.e. cannot be mapped to
each other via time or spatial translation) will depend on $m$ and
hence doing a sufficiently large number of trials will uncover the
repeated structure. It is relatively easy to see this for small
$m$. The simplest case of $m=3$, for example, has only three distinct
types of antichains, and so there are only 3 distinct possibilities
for the homology as a function of $n$. As $m$ increases, the number of
possible antichains becomes much larger, and so the test for
repetition will need more trials.

\begin{figure}[ht]
\centering \resizebox{4in}{!}{\includegraphics{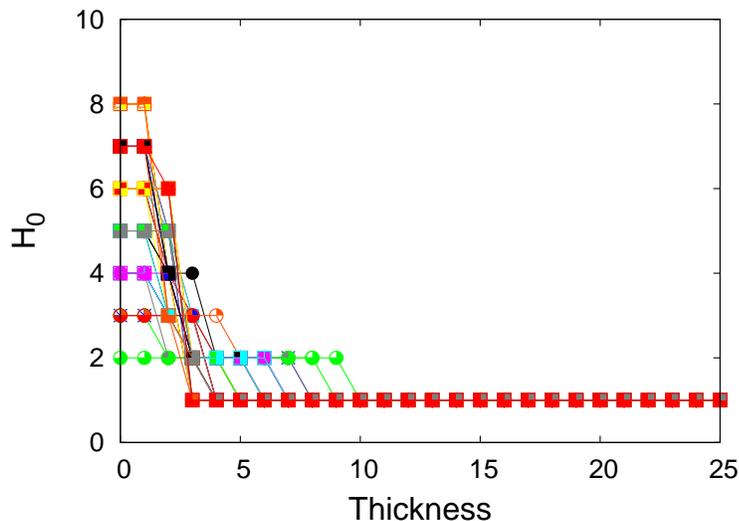}}
\caption{\small The fluctuations of the homology group $H_0$ for small
  $n$ for the $m=10$ tower of crowns in 50 trials are shown. Though
  not immediately clear from the figure, there is a repetition of the
  behaviour in a large number of the trials.}
\label{crownten}
\end{figure}

We use the random antichain algorithm to run our test on $N=1000$
element tower of crowns causal sets with $m=3$, $m=4$, $m=10$ and
$m=40$, and run the test up to the cosmological scale for the first
three and $n_m=99$ for the last.  For $m=3,4,10$, the cosmological
scale is reached early and hence no stable homology is
exhibited. Moreover, for small thickness, the homology groups are
repeated in different trials as expected, so these causal sets clearly
are not manifoldlike.  Fig \ref{crownten} shows the behaviour of
$H_0$ at small thickness for $50$ of the trials. Such perfect repetitions
signal the fact that antichains which are identical, except for being
time translated, appear more than once in the trials so that their
homology groups are identical for all $n$.

\begin{figure}[ht]
\centering \resizebox{4in}{!}{\includegraphics{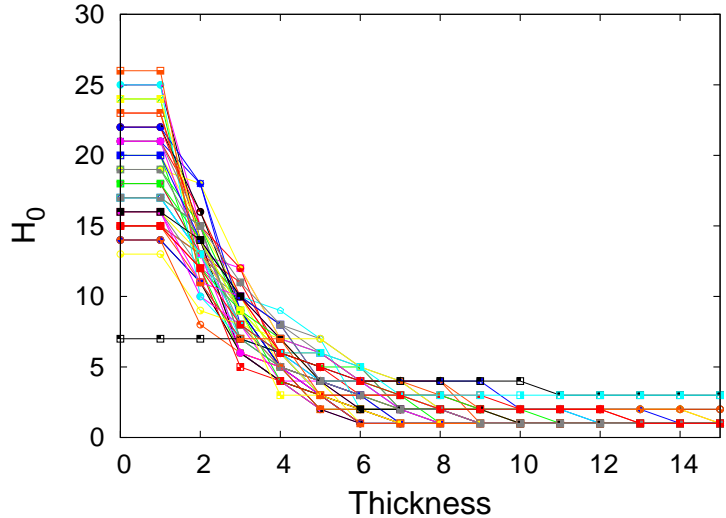}}
\caption{\small The fluctuations of the homology group $H_0$ for small
$n$ for the $m=40$ tower of crowns in $50$ trials are shown. This
  resembles more closely the fluctuations obtained from a causal set
  discretisation of the cylinder spacetime.}
\label{crownforty}
\end{figure}

For $m=40$, on the other hand, the set of $100$ trials does not
suffice to demonstrate that repetitions occur. Hence the behaviour for
small $n$ appears to have the characteristics of the stochastic
fluctuations of causal sets that are obtained from
discretisations. Fig \ref{crownforty} shows the fluctuations in
$H_0$ for $50$ trials at small thickness.  Moreover, the stability
analysis shows that for all the $93$ legitimate trials, the cylinder
homology appears as the first stable region.
Putting these ingredients together it
seems that we can conclude that the $m=40$ tower of crown causal set
satisfies our necessary but not sufficient test for manifoldlikeness
when we restrict ourselves to $100$ trials.
Having passed this topological test, however it readily fails a
crucial geometric test. The crystalline structure is immediately
obvious when counting the valency or the number of links for each
element -- it is the same (and very small) for all $1000$ elements,
and this is a 
definite sign that the causal set is not manifoldlike. While geometric
considerations rule out this simple example, it may not be as easy in
general. Importantly, this example stresses that our test is only a sufficient
condition and acts as a basic filter before using more subtle tests
for manifoldlikeness. 

We can use this example as a lesson that random antichains aren't
always sufficient and that our tests should be made more
sensitive. One could look for special classes of antichains to check
if initial fluctuations in the homology exist, are generic or the
homology versus thickening behaviour is the same for all the
antichains. As shown analytically, the tower of crowns for any $m$
will not pass this test for antichains of cardinality $m$ which
correspond to the level foliations.

\section{Conclusion} 
\label{conclusions}

In this work we have presented evidence that manifoldlikeness in a
causal set $C$ manifests itself in terms of a first stable homology
for a large fraction of randomly chosen antichains, in a statistically
large sampling of inextendible antichains.  Moreover, for small
thicknesses, the correlation between the homology across the different
antichains is very weak, this reflecting the detail of the antichain
itself, rather than being a robust feature of the continuum.

This hypothesis was reached by doing a large number of simulations for
causal sets obtained via a sprinkling into a set of 2d conformally
flat spacetimes. It was then verified for causal sets obtained from
sprinkling into a class of 3d conformally flat spacetimes.  However,
the 3d computations were computationally too intensive to allow any
statistical analysis. Nevertheless, it is a worthwhile future exercise
to demonstrate whether the above test for manifoldlikeness works for a
wider class of examples in higher dimensions. It is also important to
test how and when torsion makes its appearance. All our examples are
simple enough topologically for it not to have made an appearance, and
it would be useful to have a better understanding of this.

Our test for manifoldlikeness was then performed on examples of causal
sets not obtained via sprinklings  of continuum spacetimes. These
were causal sets generated from a class of transitive percolation
dynamics, and the regular tower of crowns causal sets. While the
former class had stable regions, these were not consistent over a
sufficient number of the trials and hence did not pass our
test. However, the restriction to relatively large values of the
parameter $p$ because of computational constraints should be borne in
mind. This means that the causal sets are very narrow, and hence not
likely to exhibit manifoldlike characteristics. Testing for
manifoldlikeness in transitive percolation dynamics for smaller values
of $p$ would therefore be worthwhile to study in the future. 

The other class of causal sets studied, the tower of crowns, do pass
the stability test, but the existence of special classes of antichains
for which there is no fluctuating homology at small thicknesses means
that they do not pass our test of manifoldlikeness.

We should also reiterate that our analytical understanding of the
homology construction is limited to globally hyperbolic regions of
spacetime. While it is true that there are large classes of spacetimes
which are not globally hyperbolic, it is also true that for strongly
causal spacetimes, there are convex normal neighbourhoods (CNNs) around
every point which are themselves globally hyperbolic.  Our algorithm
currently does not pick such neighbourhoods, but considers the global
topology of the spacetime. Ultimately, we will also need to include checks
for CNNs, for which the topology should be trivial.  For this, 
however, we would need to understand better how geometry is encoded in
a causal set.

If this test for manifoldlikeness is robust, and survives further
analysis, it would provide us a valuable tool in assessing
manifoldlikeness in causal sets generated by causal set quantum
gravity.  Conversely, it is possible that these results can suggest a
means of incorporating the right sort of locality into the quantum
dynamics \cite{Joe-Bell} so that manifoldlike causal sets emerge in
the classical limit of the theory.  

\vspace{20pt}

\noindent{{\bf Acknowledgements:}} We thank Rafael Sorkin, Sanjib
Sabhapandit and Dan Christensen for useful discussions. We are
extremely grateful to Pawel Pilarczyk for providing code to simplify
our simplicial complexes, and for numerous suggestions on how to use
his homology software (CHomP). This work was supported in part by the
Royal Society International Joint Project 2006-R2 and the computations
were carried out using the Raman Research Institute high performance
computers, and the Shared Hierarchical Academic Research Computing
Network (SHARCNET:www.sharcnet.ca).  This work was also supported in
part by the Perimeter Institute for Theoretical Physics.  Research at
Perimeter Institute is supported by the Government of Canada through
Industry Canada and by the Province of Ontario through the Ministry of
Research \& Innovation.  DR was supported in part by the Marie Curie
Research and Training Network ENRAGE (MRTN-CT-2004-005616).


\end{document}